\documentclass[reprint,NumberedRefs]{JASAnew}
\begin{document}


\title[Audiovisual joint learning for hearing aids]
{Audiovisual joint learning for end-to-end hearing aids}

\author{You-Jin Li}
\affiliation{
Graduate Institute of Communication Engineering,
National Taiwan University,
Taipei, Taiwan
}

\author{Yu Tsao}
\email{yu.tsao@citi.sinica.edu.tw}
\affiliation{
Research Center for Information Technology Innovation,
Academia Sinica,
Taipei, Taiwan
}

\author{Borching Su}
\affiliation{
Graduate Institute of Communication Engineering,
National Taiwan University,
Taipei, Taiwan
}

\author{Kuan-Chung Ting}
\affiliation{
Department of Otolaryngology-Head and Neck Surgery,
Taipei Veterans General Hospital,
Taipei, Taiwan
}
\affiliation{
School of Medicine,
National Yang Ming Chiao Tung University,
Taipei, Taiwan
}

\author{Fan-Gang Zeng}
\affiliation{
Center for Hearing Research,
University of California, Irvine,
California, USA
}

\begin{abstract}
Speech understanding in noise remains challenging for hearing-aid users, particularly in the presence of competing speakers. Conventional hearing aids typically perform speech enhancement (SE) and hearing-loss compensation in separate stages, which may cause enhancement errors and signal distortions to carry over to the amplification stage. Moreover, audio-only SE often provides limited benefits under competing-speech conditions because the target and interfering speech share similar acoustic characteristics, making them difficult to separate. To address these limitations, we propose AV-NeuroAMP, an end-to-end audiovisual framework that integrates noisy speech, target-talker video, and the listener’s audiogram to jointly perform SE, personalized amplification, and dynamic-range compression. We further introduce audiogram-conditioned feature-wise linear modulation (AC-FiLM) to effectively incorporate listener-specific hearing profiles. Objective evaluations showed that AV-NeuroAMP outperformed conventional amplification, the audio-only NeuroAMP model, and two-stage systems on an in-domain English test set, and that these improvements were retained on an unseen Mandarin test set. Listening tests involving normal-hearing participants under simulated hearing loss and listeners with hearing loss further demonstrated improvements in speech quality and intelligibility, with the greatest benefits observed under competing-speech conditions. These findings support end-to-end audiovisual personalized amplification as a promising approach for improving hearing-aid performance in challenging acoustic environments.
\end{abstract}

\maketitle
\section{Introduction}
\label{sec:intro}

More than 430 million people experience disabling hearing loss, and this number is expected to continue rising~\citep{WHO_hearing_loss_2024}. Hearing impairment is associated with communication difficulties that may lead to social isolation, depression, and reduced quality of life~\citep{shukla2020hearing,lawrence2020hearing,ciorba2012impact}. It has also been identified as a potentially modifiable risk factor for dementia in the 2024 report of the Lancet Standing Commission~\citep{livingston2024dementia,lin2013hearing}. Hearing aids remain the primary intervention for improving speech audibility~\citep{HA2012} and may alter the neural encoding of speech in older adults with hearing loss~\citep{jenkins2018amplification}. However, their benefits are often limited in the presence of background noise, reverberation, or competing speech~\citep{Gygi2016,Hohmann2023,akeroyd20232nd}, contributing to user dissatisfaction and inconsistent use~\citep{McCormack2013}. These persistent challenges, particularly those involving competing speech, motivate further advances in hearing-aid signal processing.

Most current hearing aids compensate for hearing loss through prescriptive amplification and dynamic-range compression. The National Acoustic Laboratories–Revised (NAL-R) prescription~\citep{Nal-R} specifies frequency-dependent linear gain, whereas nonlinear prescriptions, such as NAL-NL1~\citep{Nal-nl1}, NAL-NL2~\citep{Nal-nl2}, and DSL m[i/o]~\citep{dsl}, specify gain as a function of both frequency and input level. Wide dynamic range compression (WDRC)~\citep{wdrc} provides level-dependent gain to accommodate the reduced auditory dynamic range associated with hearing loss. It has also been reported that applying insertion gain at 10 kHz can successfully restore audibility for approximately 40\% of ears with mild-to-moderate hearing loss~\citep{moore2008highfrequency}.

In addition to amplification-compression, modular hearing-aid systems commonly incorporate speech enhancement (SE) to facilitate listening in noisy environments~\citep{openMHA}. Conventional single-channel SE methods include spectral subtraction~\citep{boll1979spectral,berouti1979enhancement} and statistical estimation~\citep{scalart1996wiener,mmse1,mmse2,martin2001noisepsd}; however, their performance may deteriorate under nonstationary noise and competing-speech conditions~\citep{cohen2001nonstationary,cherry1953cocktail,wang2018supervised}. Multichannel beamforming~\citep{vanveen1988beamforming,doclo2015multichannel} is another widely used approach, but its performance may also be degraded by reverberation, source or head movement, microphone mismatch, and wind noise~\citep{best2017dynamic,spriet2005robustness,zakis2011wind}. Beyond these challenges, optimizing SE and amplification-compression separately may fail to account for the final listener-specific output, allowing residual interference to be amplified along with the target speech~\citep{akeroyd20232nd,browning2019effects}. These limitations motivate approaches that jointly address SE and listener-specific compensation.

Deep learning has advanced a wide range of speech-processing applications~\citep{hinton2012deep,wang2018supervised}. It has also substantially improved SE performance, leading to the development of numerous architectures, including convolutional networks~\citep{FCN_fu2017raw,pandey2020densely}, recurrent and long short-term memory networks~\citep{RNN_valentini2016investigating,LSTM_weninger2015speech}, Transformers~\citep{vaswani2017attention}, Mamba-based models~\citep{SEMamba}, and hybrid architectures~\citep{abdulatif2024cmgan,wang2023tf}. Perceptually motivated training objectives have also been studied~\citep{fu2019metricgan,kolbaek2018monaural,fu2021metricgan+}.

Recently, deep learning has also been applied to improve hearing-aid signal processing~\citep{nwdrc2025,kayser2022spatial,dlha1,dlha2,dlha3,dlha4,ahmed2025neuroamp}. For hearing-loss compensation, auditory-model inversion trains a neural network, such that the simulated response of an impaired auditory periphery approximates a normal-hearing reference~\citep{dlha1,dlha2,dlha3}, typically using differentiable approximations of computational auditory models~\citep{dlha4}. Another approach combines learned noise reduction with controllable WDRC~\citep{nwdrc2025}. Meanwhile, a neural-network-based amplification system, termed NeuroAMP, was proposed to learn a direct mapping from an acoustic input and a listener’s audiogram to a hearing-loss-compensated output~\citep{ahmed2025neuroamp}. However, these hearing-loss compensation approaches do not incorporate visual cues. Audio-only processing may still struggle under strong interference and competing-speech conditions, where it may fail to fully separate the target speech or introduce distortions that degrade perceived quality~\citep{brons2014effects,kolbaek2016speech}. In this study, we extend neural compensation by incorporating visual cues to address these remaining challenges.

Auditory and visual cues provide partially independent yet complementary information. Visual signals convey articulatory configurations and place-of-articulation information, including viseme-related cues~\citep{owens1985visemes}, whereas auditory temporal-envelope and spectral cues convey consonant distinctions that may be visually ambiguous~\citep{vantasell1987envelope}. The frequency-band importance function for consonant recognition also changes when visual cues are available, indicating that visual information complements rather than duplicates acoustic information in specific frequency bands~\citep{bernstein2020bandimportance}. The audiovisual benefit observed in listeners with hearing loss has been attributed to these complementary cues~\citep{grant1998avintegration}. Seeing a talker’s face also helps children with hearing loss recognize speech in noise~\citep{lalonde2022face}.

Complementary visual information has also been leveraged to develop deep learning–based audiovisual speech-enhancement (AVSE) models. Numerous studies have demonstrated that incorporating visual cues enables AVSE models to outperform audio-only SE systems, particularly at low signal-to-noise ratios (SNRs) and under competing-talker conditions~\citep{hou2018audio,ephrat2018looking,michelsanti2021overview,AVSE5,Conversion,yang2026tracking}. A diverse range of deep-learning architectures has been explored for AVSE, ranging from convolutional and recurrent fusion networks~\citep{michelsanti2021overview,hou2018audio,TMSV} to Transformer-based~\citep{AVTransformer2023} and Mamba-based~\citep{SAVSE2025} models. To facilitate efficient online processing, lightweight architectures have also been developed to reduce computational complexity~\citep{martel2023lightweight}.

In this study, we extend NeuroAMP~\citep{ahmed2025neuroamp}, which demonstrated the potential benefits of jointly optimizing SE and neural hearing-loss compensation through end-to-end training, by incorporating visual cues into the proposed AV-NeuroAMP framework. AV-NeuroAMP jointly performs AVSE and listener-specific hearing-loss compensation. We further introduce audiogram-conditioned feature-wise linear modulation (AC-FiLM) to effectively integrate listener-specific audiogram information into the framework.

To evaluate AV-NeuroAMP, we conducted comprehensive objective and subjective experiments, along with band-specific gain analyses and an ablation study assessing the effectiveness of AC-FiLM. The results demonstrate that AV-NeuroAMP outperforms competing baselines, including conventional hearing-loss compensation, the audio-only NeuroAMP model, and two-stage SE–compensation systems in objective and subjective evaluations under noisy conditions. They also show that AC-FiLM outperforms direct audiogram concatenation within the same audiovisual backbone. Finally, AV-NeuroAMP demonstrates promising cross-lingual generalization.

The main contributions of this work are fourfold: (1) We propose AV-NeuroAMP, the first framework to integrate an audiovisual front end with neural hearing-loss compensation, achieving promising performance under challenging noisy conditions. (2) We develop AC-FiLM, which effectively incorporates listener-specific audiogram information into the neural amplification framework through frequency-dependent feature modulation. (3) Through objective evaluations, ablation studies, and band-specific gain analyses, we demonstrate that AV-NeuroAMP outperforms competing baselines and exhibits promising cross-lingual generalization. (4) We validate the perceptual benefits of AV-NeuroAMP through listening tests involving both listeners with hearing loss and normal-hearing listeners under simulated hearing loss. The results demonstrate improved speech quality and intelligibility, particularly under competing-speech conditions.

Collectively, these contributions highlight the potential of integrating multimodal information with neural hearing-loss compensation to advance the design of next-generation hearing aids.



\section{Methods}
\label{sec:methods}

Given an input speech signal \(\mathbf{y}\), consisting of target speech \(\mathbf{s}\), background noise \(\mathbf{b}\), and interfering speech \(\mathbf{i}\), such that \(\mathbf{y}=\mathbf{s}+\mathbf{b}+\mathbf{i}\), a hearing-aid system processes \(\mathbf{y}\) together with the listener’s audiogram \(\mathbf{a}\) to produce a listener-specific amplified output \(\mathbf{s}_{\mathrm{Amp}}\). Conventional hearing-aid systems typically adopt a two-stage architecture, in which an SE front end first enhances \(\mathbf{y}\), followed by a hearing-loss compensation stage that generates the final output \(\mathbf{s}_{\mathrm{Amp}}\) based on the listener’s audiogram \(\mathbf{a}\).

NeuroAMP employs an end-to-end neural network to directly estimate \(\mathbf{s}_{\mathrm{Amp}}\) from \(\mathbf{y}\) and \(\mathbf{a}\), with the training target generated by applying NAL-R and WDRC to the clean target speech. AV-NeuroAMP further incorporates visual cues \(\mathbf{v}\) to jointly perform SE and personalized hearing-loss compensation. To facilitate neural hearing-loss compensation, each listener is represented by an audiogram vector containing hearing thresholds at six standard audiometric frequencies.

\begin{equation}
\label{eq:audiogram}
\mathbf{a} = [a_{250},a_{500},a_{1000},a_{2000},a_{4000},a_{6000}] \in\mathbb{R}^{6},
\end{equation}

where each element is expressed in decibels hearing level (dB HL). To facilitate model training, the audiogram values are scaled by a fixed constant \(C_A\), such that \(\widetilde{\mathbf{a}}=\mathbf{a}/C_A\). This scaling operation is referred to as \texttt{LevelNorm} in the following discussion.

\begin{figure*}[ht]
\begin{center}
\includegraphics[width =1.8\reprintcolumnwidth]{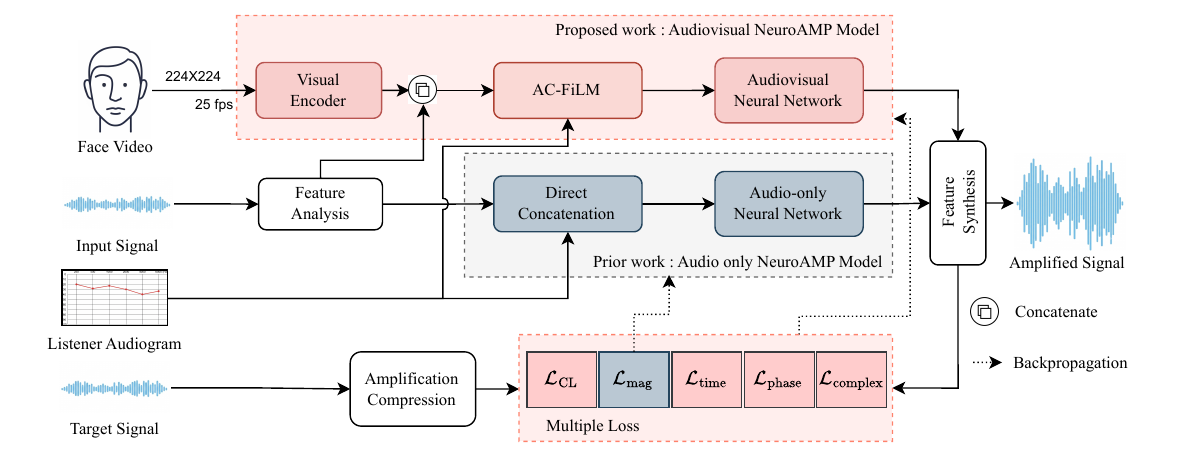}
\caption{\label{fig:arch}{Overview of NeuroAMP~\citep{ahmed2025neuroamp} and AV-NeuroAMP. The gray regions denote the audio-only NeuroAMP pathway with a direct audiogram concatenation and a spectral MSE loss. The pink regions denote AV-NeuroAMP, which includes a visual encoder, AC-FiLM, and a multi-loss function. Unfilled blocks represent shared operations or inputs, including speech analysis, speech synthesis, and the amplification and compression process. Dotted arrows indicate the training backpropagation paths.}}
\end{center}
\end{figure*}

Figure~\ref{fig:arch} compares NeuroAMP with AV-NeuroAMP (Sec.~\ref{subsec:proposed}). The gray region represents the audio-only NeuroAMP pathway, whereas the pink region represents AV-NeuroAMP. The main extensions from NeuroAMP to AV-NeuroAMP include the incorporation of a visual input, AC-FiLM, and a multi-objective loss function. In the following section, we first describe NeuroAMP and then detail the proposed AV-NeuroAMP framework in the subsequent section.

\subsection{NeuroAMP: audio-only neural amplification}
\label{subsec:prior}

NeuroAMP~\citep{ahmed2025neuroamp} directly maps noisy speech and a listener-specific audiogram to personalized amplified speech. The input speech \(\mathbf{y}\) is transformed into spectral features \(\mathbf{Y}\) through a speech analysis operation. The mapping of the audiogram encoder, \(\mathcal{F}_{\mathrm{audiogram}}\), converts the six-dimensional normalized audiogram vector, \(\widetilde{\mathbf{a}}\), to generate an audiogram representation $\mathbf{H}_{\mathrm{A}}$, a representation that is compatible with the spectral features for direct concatenation. In~\citep{ahmed2025neuroamp}, the audiogram encoder is implemented as a neural network comprising two linear layers with a SiLU activation between them.

NeuroAMP then estimates the amplified spectral features from the concatenated representation of \(\mathbf{Y}\) and \(\mathbf{H}_{\mathrm{A}}\):

\begin{equation}
\label{eq:neuroamp}
\hat{\mathbf{S}}^{\mathrm{NeuroAMP}}_{\mathrm{Amp}}
= \mathcal{F}_{\mathrm{NeuroAMP}}(\left[\mathbf{Y} \Vert \mathbf{H}_{\mathrm{A}}\right]).
\end{equation}
where \(\left[\cdot \,\Vert\, \cdot\right]\) denotes concatenation along the channel dimension, and \(\mathcal{F}_{\mathrm{NeuroAMP}}\) denotes the overall mapping performed by the NeuroAMP model. The model was trained using a spectral-domain mean squared error (MSE) loss between the predicted amplified spectrum and the corresponding ground-truth spectrum. During inference, the input speech is first transformed into spectral features through the speech analysis operation and then processed by \(\mathcal{F}_{\mathrm{NeuroAMP}}\). The resulting spectral features, \(\hat{\mathbf{S}}_{\mathrm{Amp}}^{\mathrm{NeuroAMP}}\), are then converted into the time-domain amplified speech signal, \(\hat{\mathbf{s}}_{\mathrm{Amp}}^{\mathrm{NeuroAMP}}\), through a speech synthesis operation. In NeuroAMP~\citep{ahmed2025neuroamp} (and AV-NeuroAMP in this study), the short-time Fourier transform (STFT) and inverse STFT (iSTFT) are used as the speech analysis and synthesis operations, respectively.

NeuroAMP has been implemented using convolutional, long short-term memory, convolutional recurrent, and Transformer architectures~\citep{ahmed2025neuroamp}. In this study, we adopted the Transformer variant, which achieved the best performance in~\citep{ahmed2025neuroamp}, as the audio-only NeuroAMP baseline (Sec.~\ref{subsec:comparison}).

\subsection{AV-NeuroAMP: audiovisual neural amplification}
\label{subsec:proposed}

The overall architecture of AV-NeuroAMP is illustrated by the pink regions in Fig.~\ref{fig:arch}. Compared with NeuroAMP, AV-NeuroAMP incorporates a visual encoder that extracts a visual representation, \(\mathbf{H}_{\mathrm{V}}\), from the visual input and combines it with the spectral features, \(\mathbf{Y}\). The AC-FiLM module then conditions the fused features on the listener’s audiogram, and a multi-objective loss function is used to train the overall model.

\subsubsection{Visual feature extraction}
\label{subsec:visual}

\begin{figure}[ht]
\begin{center}
\includegraphics[width=\columnwidth]{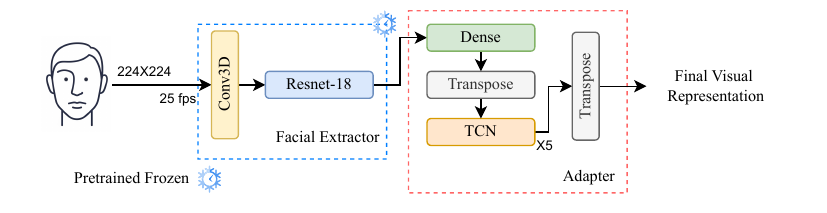}
\caption{\label{fig:visualenc}{Diagram of AV-NeuroAMP visual encoder: a frozen pretrained facial extractor followed by an adapter module whose output is aligned with the audio features.}}
\end{center}
\end{figure}

The visual encoder in AV-NeuroAMP comprises two components: a facial extractor module and an adapter module (Fig.~\ref{fig:visualenc}). Given the input video sequence, \(\mathbf{V}=[\mathbf{v}_1,\ldots,\mathbf{v}_{T_{\mathrm{V}}}]\), comprising \(T_{\mathrm{V}}\) frames, the facial extractor module applies a three-dimensional convolution followed by an 18-layer residual network to extract a face-only representation, $\mathbf{H}_{\mathrm{face}}$, where   $\mathbf{H}_{\mathrm{face}} = \mathcal{F}_{\mathrm{ResNet\text{-}18}} \left(\operatorname{Conv3D}(\mathbf{V})\right). $ 
The adapter module, whose mapping is indicated by \(\mathcal{F}_{\mathrm{VC}}\), is implemented as a temporal convolutional network (TCN) comprising a linear projection followed by five residual one-dimensional convolutional blocks. By processing \(\mathbf{H}_{\mathrm{face}}\) through the adapter module, the final visual representation, $\mathbf{H}_{\mathrm{V}}$, is obtained via: $\mathbf{H}_{\mathrm{V}} = \mathcal{F}_{\mathrm{VC}}\left(\mathbf{H}_{\mathrm{face}}\right)$.

Because training a robust facial feature extractor requires a large amount of data, limited training data may result in biased representations. Therefore, we pretrained the facial feature extractor on the Second Audio-Visual Speech Enhancement Challenge (AVSEC-2) dataset~\citep{10023284} and froze its parameters during AV-NeuroAMP training. In contrast, the TCN-based adapter module was jointly trained with the other components of AV-NeuroAMP.

\subsubsection{Audiogram-conditioned feature-wise linear modulation (AC-FiLM)}
\label{subsec:acfilm}
In NeuroAMP, the audiogram representation is directly concatenated with the audio features, as described in Sec.~\ref{subsec:prior}. In AV-NeuroAMP, the visual representation \(\mathbf{H}_{\mathrm{V}}\) (Sec.~\ref{subsec:visual}) is additionally combined with the audio features to form an audiovisual representation. When direct concatenation is used to incorporate the audiogram representation, AV-NeuroAMP generates the final compensated spectral features as

\begin{equation}
\label{eq:AVneuroamp}
\hat{\mathbf{S}}^{\mathrm{AV-NeuroAMP}}_{\mathrm{Amp}}
= \mathcal{F}_{\mathrm{AV-NeuroAMP}}(\left[\mathbf{Y} \Vert \mathbf{H}_{\mathrm{V}} \Vert \mathbf{H}_{\mathrm{A}}\right]).
\end{equation}

where \(\mathcal{F}_{\mathrm{AV-NeuroAMP}}\) denotes the overall mapping performed by the AV-NeuroAMP model.

Although direct concatenation offers a straightforward and previously validated approach to incorporating an audiogram into an audio representation, it has inherent limitations. An audiogram is a low-dimensional, time-invariant profile, whereas an audiovisual representation is high-dimensional and temporally varying. When the fused features are subsequently processed using instance normalization~\citep{ulyanov2016instance}, their channel-wise means are removed and variances are normalized. This operation may weaken listener-specific information encoded in absolute feature offsets or scales, making audiograms with similar frequency-dependent patterns but different overall degrees of hearing loss less distinguishable. This structural limitation of applying audiogram conditioning before normalization highlights the need for careful design of end-to-end hearing-loss compensation frameworks.

Recently, feature-wise linear modulation (FiLM) has emerged as a more explicit mechanism for conditional feature adaptation~\citep{Perez,birnbaum2019temporal,MeseguerBrocal2019}. Rather than directly concatenating a conditioning vector with the input, FiLM transforms the vector into feature-wise scaling and shifting parameters that modulate an intermediate representation. The scaling parameters enable the conditioning information to emphasize or suppress specific features, whereas the shifting parameters introduce condition-dependent offsets. FiLM therefore establishes both multiplicative and additive interactions between the conditioning information and the target representation. Its effectiveness as an alternative to concatenation-based conditioning has been demonstrated in various audio-processing models~\citep{yeh2024hyperrnn,ni2026affine}.

Building on this principle, we propose audiogram-conditioned feature-wise linear modulation (AC-FiLM), which incorporates listener-specific hearing information into the audiovisual representation after instance normalization. AC-FiLM extends conventional FiLM in three key respects. First, it employs complementary local- and global-conditioning branches to capture frequency-specific variations in hearing thresholds and listener-level characteristics shared across frequency regions, respectively. Second, it generates channel- and frequency-dependent scaling and shifting parameters, enabling fine-grained modulation of the audiovisual representation. Third, AC-FiLM introduces a gated residual connection that adaptively controls the strength of audiogram conditioning while preserving the original audiovisual features. This design follows the general principle of gated residual conditioning adopted in adaLN-Zero, in which condition-dependent scaling, shifting, and gating parameters are incorporated into a zero-initialized residual block~\citep{peebles2023dit}. Unlike adaLN-Zero, however, AC-FiLM employs a sigmoid-activated gate and derives its modulation parameters specifically from the listener’s audiogram. Applying these parameters after instance normalization prevents the audiogram information from being normalized together with the audiovisual features, thereby providing a more explicit mechanism for preserving both frequency-specific hearing-loss patterns and information about the listener’s overall hearing-loss level.

\begin{figure}[ht]
\begin{center}
\includegraphics[width=\columnwidth]{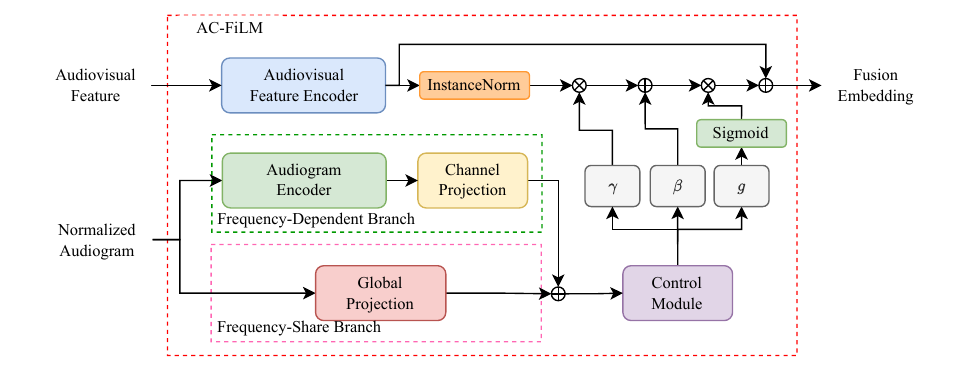}
\caption{\label{fig:acfilm}{Overview of the proposed AC-FiLM module. The normalized audiogram is processed by local-conditioning and global-conditioning branches to produce complementary frequency-varying and frequency-shared conditions. Their fused representation is mapped by the control module to generate the scaling parameter $\boldsymbol{\gamma}$, shifting parameter $\boldsymbol{\beta}$, and gating parameter $\mathbf{g}$. These parameters modulate the instance-normalized audiovisual features, while the residual connection preserves the encoded audiovisual representation, producing the listener-specific fusion embedding $\mathbf{H}_{\mathrm{fusion}}$.}}
\end{center}
\end{figure}

Figure~\ref{fig:acfilm} illustrates the overall architecture of the proposed AC-FiLM module. First, the audio and visual features are concatenated to form $\left[\mathbf{Y}\Vert\mathbf{H}_{\mathrm{V}}\right]$ and are subsequently processed by the audiovisual feature encoder to produce $\mathbf{H}_{\mathrm{AV}}$. This operation maps the input time--frequency representation into a higher-dimensional audiovisual feature space comprising channel, time, and frequency dimensions.

In parallel, the normalized audiogram vector $\widetilde{\mathbf{a}}$ is processed by two complementary conditioning branches. The local-conditioning branch generates an audiogram representation that varies across frequency bins and feature channels. The global-conditioning branch produces a channel-dependent representation that is shared across all frequency bins, providing a common listener-level condition at each frequency position. Although both branches receive the complete audiogram, they produce complementary conditioning features: the local-conditioning features vary across frequency bins, whereas the global-conditioning features are shared across all frequency bins. These conditions are subsequently fused and mapped to the scaling, shifting, and gating parameters used to modulate the encoded audiovisual representation. The resulting representation, $\mathbf{H}_{\mathrm{fusion}}$, therefore incorporates listener-specific audiogram information without directly concatenating the audiogram with the audiovisual features.

The local-conditioning branch, \(\mathcal{F}_{\mathrm{FD}}\), comprises the same audiogram encoder used in the direct-concatenation method, followed by a channel-mapping layer implemented as a single linear layer. This branch transforms the six-dimensional normalized audiogram vector, \(\widetilde{\mathbf{a}}\), into a frequency-dependent conditioning representation, \(\mathbf{H}_{\mathrm{fd}}\), where
$\mathbf{H}_{\mathrm{fd}} = \mathcal{F}_{\mathrm{FD}} \left( \widetilde{\mathbf{a}} \right)$. The dimensions of \(\mathbf{H}_{\mathrm{fd}}\) are matched to those of \(\mathbf{H}_{\mathrm{AV}}\), enabling the audiogram-derived condition to be aligned with the corresponding frequency bins and feature channels.

The global-conditioning branch \(\mathcal{F}_{\mathrm{FS}}\) uses a single linear layer to project the normalized audiogram vector \(\widetilde{\mathbf{a}}\) to a \(C\)-dimensional vector, where \(C\) is the number of channels in the encoded audiovisual representation. This vector is then broadcast across the \(F'\) frequency bins to form the global conditioning feature \(\mathbf{H}_{\mathrm{fs}}\in\mathbb{R}^{C\times F'}\). Thus, each frequency bin receives the same channel-dependent audiogram representation. The complete mapping, including the linear projection and broadcasting, is expressed as \(\mathbf{H}_{\mathrm{fs}}=\mathcal{F}_{\mathrm{FS}}(\widetilde{\mathbf{a}})\).

Then, the global-conditioning feature\(\mathbf{H}_{\mathrm{fs}}\) is fused with the local-conditioning feature \(\mathbf{H}_{\mathrm{fd}}\) to form a unified audiogram conditioning feature \(\mathbf{H}_{\mathrm{AC}}\) containing both global characteristics and frequency-specific details.

\begin{equation}
\mathbf{H}_{\mathrm{AC}} = \mathbf{H}_{\mathrm{fd}} + \mathbf{H}_{\mathrm{fs}}.
\end{equation}

To convert the unified audiogram conditioning feature into the modulation parameters, we employ a control module $\mathcal{F}_{CM}$ consisting of a SiLU activation followed by a linear layer. The module maps the unified audiogram conditioning feature, \(\mathbf{H}_{\mathrm{AC}}\), to three outputs of the same size, corresponding to the scaling parameter \(\boldsymbol{\gamma}\), shifting parameter \(\boldsymbol{\beta}\), and gating parameter \(\mathbf{g}\), respectively. The operation is expressed as

\begin{equation}
\left[ \boldsymbol{\gamma} \mathbin{\Vert} \boldsymbol{\beta} \mathbin{\Vert} \mathbf{g} \right] = \mathcal{F}_{\mathrm{CM}} \left(\mathbf{H}_{\mathrm{AC}}\right)\in \mathbb{R}^{3C\times F'},
\end{equation}

Finally, the learned scaling $\boldsymbol{\gamma}$, shifting $\boldsymbol{\beta}$, and gating $\mathbf{g}$ parameters are applied to the audiovisual encoded representation \(\mathbf{H}_{\mathrm{AV}}\). Because the audiogram remains fixed throughout an utterance, these parameters are broadcast along the time dimension, providing consistent listener-specific conditioning across all time frames. The AC-FiLM modulation is formulated as follows:

\begin{equation}
\label{eq:acfilm}
\mathbf{H}_{\mathrm{fusion}} = \mathbf{H}_{\mathrm{AV}} + \sigma(\mathbf{g}) \odot \left[ \boldsymbol{\gamma} \odot \operatorname{IN}(\mathbf{H}_{\mathrm{AV}}) + \boldsymbol{\beta} \right],
\end{equation}

where $\operatorname{IN}(\cdot)$ denotes instance normalization, $\sigma(\cdot)$ is the sigmoid function, and $\odot$ represents element-wise multiplication. Before modulation, $\boldsymbol{\gamma}$, $\boldsymbol{\beta}$, and $\mathbf{g}$ are expanded along the time dimension and shared across all time frames. The scaling parameter $\boldsymbol{\gamma}$ adjusts the response of each channel frequency position, while the shifting parameter $\boldsymbol{\beta}$ introduces an audiogram-dependent offset after normalization. The gating parameter $\mathbf{g}$ adaptively controls the contribution of the conditioned features through $\sigma(\mathbf{g})$.

The residual connection in Eq.~\eqref{eq:acfilm} preserves the original audiovisual representation while introducing listener-specific modifications. This prevents the audiogram-conditioned modulation from overwriting the audio and visual information contained in $\mathbf{H}_{\mathrm{AV}}$. In addition, the output layer of the control module is initialized to zero. Consequently, the modulation term is initially zero and $\mathbf{H}_{\mathrm{fusion}}$ approximates $\mathbf{H}_{\mathrm{AV}}$ at the beginning of training. The model can therefore gradually learn the required listener-specific modulation without disrupting the original audiovisual representation.

Taken together, AC-FiLM provides an explicit alternative to direct audiogram concatenation. The local-conditioning branch produces the frequency-dependent feature \(\mathbf{H}_{\mathrm{fd}}\), which represents variations in hearing thresholds across frequency regions, whereas the global-conditioning branch produces the frequency-shared feature \(\mathbf{H}_{\mathrm{fs}}\), which provides a listener-level summary derived from the entire audiogram. Their combination, \(\mathbf{H}_{\mathrm{AC}}\), serves as the conditioning signal from which the scaling, shifting, and gating parameters are generated. Because these parameters are applied after instance normalization, the audiogram information is not directly normalized together with the audiovisual representation. The resulting feature, \(\mathbf{H}_{\mathrm{fusion}}\), therefore retains the original audiovisual information while incorporating both the frequency-dependent pattern and the overall severity of the listener's hearing loss. This listener-specific audiovisual representation is subsequently passed to the decoders to generate the personalized amplified speech.

\subsubsection{Training target and loss function}
\label{subsec:loss}

The proposed AV-NeuroAMP estimates a personalized SE and amplification target, \(\mathbf{S}_{\mathrm{Amp}}\), generated by applying NAL-R and WDRC to clean target speech. In this study, we implemented the audiovisual neural network in Fig. \ref{fig:arch} based on the state-of-the-art: AVMamba~\citep{SAVSE2025}. Because AVMamba employs a dual-path architecture to estimate both the magnitude and phase of the target speech, the model is trained using a combined loss, \(\mathcal{L}_{\mathrm{Total}}\), formulated as follows:

\begin{equation}
\mathcal{L}_{\mathrm{Total}} =
c_1\mathcal{L}_{\mathrm{CL}}
+c_2\mathcal{L}_{\mathrm{time}}
+c_3\mathcal{L}_{\mathrm{mag}}
+c_4\mathcal{L}_{\mathrm{complex}}
+c_5\mathcal{L}_{\mathrm{phase}},
\end{equation}

where \(\mathcal{L}_{\mathrm{CL}}\), \(\mathcal{L}_{\mathrm{time}}\), \(\mathcal{L}_{\mathrm{mag}}\), \(\mathcal{L}_{\mathrm{complex}}\), and \(\mathcal{L}_{\mathrm{phase}}\) denote the consistency~\citep{CL}, time-domain, magnitude, complex-spectral, and phase losses, respectively. The corresponding loss coefficients are provided in Sec.~\ref{sec:Model_Implementation} and follow the configuration reported in~\citep{SAVSE2025,SEMamba}.

\subsection{Audiovisual datasets and audiograms}
\label{subsec:data}

In this study, we used listener-specific hearing-loss data from the Clarity Challenge~\citep{graetzer2021clarity}. For the audiovisual data, we used AVSEC-2~\citep{10023284} for model training and in-domain evaluation, and the Taiwan Mandarin Speech with Video (TMSV) corpus~\citep{TMSV} for cross-lingual out-of-domain evaluation and subjective listening tests.

\subsubsection{Clarity audiograms}
The Clarity Challenge provides 272 ear-specific audiograms: 166 for training and 106 for testing~\citep{graetzer2021clarity}. Each contains air-conduction thresholds at 250, 500, 1000, 2000, 3000, 4000, 6000, and 8000~Hz, covering the conventional pure-tone audiometric range~\citep{ASHA2005}. The six thresholds at 250, 500, 1000, 2000, 4000, and 6000~Hz were used as model inputs and for the NAL-R prescription.

\subsubsection{AVSEC-2: training and in-domain evaluation}
\label{subsubsec:AVSEC-2}
AVSEC-2, constructed from the Lip Reading Sentences 3 (LRS3) corpus~\citep{afouras2018lrs3}, contains synchronized noisy audiovisual scenes and their clean-speech references. Interference includes competing speech from LRS3 and environmental noise from the Clarity~\citep{graetzer2021clarity}, DEMAND~\citep{thiemann2013diverse}, and Deep Noise Suppression Challenge~\citep{reddy2021icassp} datasets.

The training set contains 34,524 audiovisual scene pairs, including clean speech, noisy speech, and face visual. These scenes were generated from 605 target speakers, 405 competing speakers, and 7,346 noisy recordings, covering 15 categories. Four audio samples were randomly selected from 166 training audiograms for each scene, resulting in 138,096 audiogram-conditioned training pairs via the NAL-R/WDRC process. The development set, used for in-domain evaluation according to the official AVSEC-2 protocol, contains 3,306 scenes from 85 target speakers, 30 competing speakers, and 1,825 noisy recordings; the target speakers were not present in the training set. One audiogram sample was randomly selected from 106 test audiograms for each scene, generating 3,306 personalized evaluation pairs.

\subsubsection{TMSV: out-of-domain evaluation and subjective listening tests}
A subset of TMSV~\citep{TMSV} was used for cross-lingual objective evaluation and listening tests. It contains 240 phonetically balanced Mandarin utterances from one male and one female talker, each contributing 120 ten-character sentences lasting approximately 2--4~s. For the present study, noisy utterances were generated with background talkers, pink noise, babble noise, engine noise, and street noise at SNRs of $8$, $5$, $2$, $-1$, and $-4$~dB, yielding 6,000 noisy samples. For objective evaluation, each utterance was assigned one of the 106 test audiograms to generate its NAL-R/WDRC target. In the listening experiment with listeners with hearing loss, targets were generated from the audiogram of the tested ear.

\subsection{Competing baseline systems}
\label{subsec:comparison}

In this study, four competing baseline systems and the proposed system were compared: 

1. \textbf{Conventional}: A system that directly amplifies the noisy input using NAL-R/WDRC personalized amplification $Y_Amp$. 

2. \textbf{Two-stage(A)}: Two-stage audio-only SE and personalized amplification, in which SEMamba~\citep{SEMamba} is finetuned on the training set described in Section~\ref{subsubsec:AVSEC-2} to serve as the audio-only SE processor, followed by NAL-R/WDRC-based personalized amplification of the enhanced output.

3. \textbf{Two-stage(AV)}: A system configured similarly to the \textbf{Two-stage (A)} system, in which AVMamba~\citep{SAVSE2025} is fine-tuned on the training set described in Section~\ref{subsubsec:AVSEC-2} to serve as the audiovisual SE processor, followed by NAL-R/WDRC-based personalized amplification of the enhanced output. 

4. \textbf{NeuroAMP}: An audio-only end-to-end baseline based on the Transformer variant of NeuroAMP, which achieved the best performance in~\citep{ahmed2025neuroamp}. The model is trained from scratch using the in-domain training set described in Section~\ref{subsubsec:AVSEC-2}.

5. \textbf{AV-NeuroAMP}: The proposed AV-NeuroAMP methods. 

In this study, the implementation and processing settings of NAL-R/WDRC are based on the Clarity Challenge GitHub code repository \footnote{\url{https://github.com/claritychallenge/clarity}}.

\subsection{Model implementation details}
\label{sec:Model_Implementation}

This section details the implementation of the proposed AV-NeuroAMP model. Audio samples are taken at a frequency of 16 kHz and analyzed using a 512-point STFT. A periodic Hann window is employed, with 128 samples hopping to obtain 257 frequency intervals. Power-law compression with an exponent of 0.3 is used for amplitude. The video is a 25 frames/second, 224×224 pixel full-face video, converted to grayscale and scaled to the [0,1] interval. During training, a synchronized 2-second segment (32,000 audio samples and 50 frames of video) is randomly cropped from the audio and video, with shorter samples zero-padding.

Except for the AC-FiLM module, the parameter settings of the AV-NeuroAMP model are largely consistent with AVMamba~\citep{SAVSE2025}. After feature extraction, the signal is processed through eight time-frequency Mamba modules, followed by a magnitude and phase decoder to obtain the estimated signal. The AV-NeuroAMP model was trained on four NVIDIA H100 GPUs for 200 epochs with a global batch size of 16. The AdamW optimizer was used for training, with an initial learning rate of $5 \times 10^{-4}$, $\beta_1 = 0.8$, $\beta_2 = 0.99$, and an exponential learning rate decay factor of $0.99$ per epoch. The combined loss coefficients $\left[c_1, c_2, c_3, c_4, c_5 \right] = \left[0.1, 0.2, 0.9, 0.1, 0.3\right]$. During the inference phase, complete speech was processed without segmentation, and the output waveform matched the original length.

The AV-NeuroAMP model contains 18.019 million parameters, comprising 6.834 million trainable parameters and 11.185 million frozen parameters in the visual front end. Table~\ref{tab:complexity} presents a module-wise complexity analysis. The visual encoder accounts for the largest share of the parameters, whereas the audiovisual neural network (AVNN) accounts for the largest share of the computational cost (137.38 GMACs), followed by the visual encoder (94.78 GMACs). In contrast, AC-FiLM contains only 21,824 trainable parameters and requires fewer than 0.01 GMACs; this overhead is negligible relative to the 232.16 GMACs required to process a two-second audiovisual input.

\begin{table}[t]
\centering
\renewcommand{\arraystretch}{1.15}
\caption{Parameter count and computational cost of AV-NeuroAMP by module, for a 2-s audiovisual input. The frozen facial extractor module is used at inference and is therefore included in the GMAC count. Parameter counts are rounded to 0.001~M; parameter totals are the sums of the displayed module values.}
\label{tab:complexity}
\begin{tabular}{lrrr}
\hline
\textbf{Module} & \textbf{Parameters} & \textbf{Trainable} & \textbf{GMAC} \\
\hline
Visual encoder            & 14.627~M & 3.442~M & 94.78 \\
AVNN                      & 3.370~M  & 3.370~M & 137.38 \\
AC-FiLM conditioner       & 0.022~M  & 0.022~M & $<0.01$ \\
\hline
Total                     & 18.019~M & 6.834~M & 232.16 \\
\hline
\end{tabular}
\end{table}

\subsection{Objective evaluation metrics}
\label{sec:objective_metrics}

To evaluate the effectiveness of the proposed AV-NeuroAMP, the Hearing Aid Speech Perception Index (HASPI, version 2)~\citep{Haspi1,Haspi2} and the Hearing Aid Speech Quality Index (HASQI, version 2)~\citep{Hasqi1} were used to assess the intelligibility and quality, respectively, of the personalized amplified speech. Audiometric thresholds at 250, 500, 1000, 2000, 4000, and 6000 Hz were used in the evaluation. MSE was employed to measure waveform-level consistency between the model output and the target personalized amplified speech, while the signal-to-noise ratio (SNR) was used to evaluate the overall noise-reduction capability.

To verify whether the proposed method can jointly perform SE and personalized amplification across all frequency bands, we introduce the Multi-Band Short-Time Gain Error (MBGE), a new metric that measures the consistency between the model output and the target in terms of short-time, band-specific signal levels. Spectral analysis is performed using a 512-point STFT with a Hann window and a hop size of 128 samples. The power-law-compressed magnitude is first restored to the linear domain using the inverse compression factor. The frequency range is then divided into six non-overlapping bands centered at 250, 500, 1000, 2000, 4000, and 6000 Hz. Let \(R_{f,t}(\mathbf{Z})\) denote the root-mean-square magnitude of spectrum \(\mathbf{Z}\) within frequency band \(f\) at frame \(t\). The short-time gain error is defined as follows:

\begin{equation}
\label{eq:gainerr}
e_{f,t}^{\mathrm{gain}} = \left| 20\log_{10} \left( \frac{ R_{f,t}(\hat{\mathbf{S}}_{\mathrm{Amp}})+\epsilon }{ R_{f,t}(\mathbf{S}_{\mathrm{Amp}})+\epsilon } \right) \right|,
\end{equation}

where $\hat{\mathbf{S}}_{\mathrm{Amp}}$ and $\mathbf{S}_{\mathrm{Amp}}$ are the model output and the corresponding NAL-R/WDRC target signal, respectively. Voice activity detection (VAD) is performed using the RMS values of the six frequency bands of the clean signal $\mathbf{S}$ to obtain the active frames $\mathcal{A}_{f}$ for each frequency band. Then, the MBGE of each band is calculated by averaging the effective ranges of:

\begin{equation}
\label{eq:MBGE_f}
\mathrm{MBGE}_f = \frac{1}{\mathcal{A}_{f}}
\sum_{t\in\mathcal{A}_{f}} e_{f,t}^{\mathrm{gain}}
\end{equation}

Finally, the average MBGE per band can be obtained as the total MBGE over all bands:

\begin{equation}
\label{eq:MBGE}
\mathrm{MBGE} = \frac{1}{6}
\sum_{f\in6} \mathrm{MBGE}_f 
\end{equation}

An MBGE score is computed for each speech input, and the table reports the mean and standard deviation across all inputs under each condition. A lower MBGE indicates greater consistency between the model output and the target in terms of short-time, band-specific signal levels. For noisy inputs, the error may reflect both residual interference and deviations from the prescribed amplification gain. We believe that this new metric provides a more precise measure of the differences between the target amplified speech and the model output across time and frequency bands.

\subsection{Audiogram-conditioning ablation study}
\label{subsec:ablationdesign}

To demonstrate the effectiveness of the proposed AC-FiLM, an ablation experiment compared the objective evaluation scores of two different audiogram fusion methods within the same audiovisual backbone network: 1. direct concatenation, and 2. AC-FiLM. Both methods utilize the same training data, listener-specific training objectives, loss function, optimization settings, and evaluation protocol, differing only in the audiogram encoding and concatenation fusion methods.

\subsection{Listening tests}
\label{subsec:listening}

To demonstrate the clinical effectiveness of the proposed AV-NeuroAMP method, we conducted two listening experiments: (1) a subjective evaluation involving normal-hearing listeners under simulated hearing-loss conditions and (2) an evaluation involving listeners with hearing loss. Before describing the experimental procedures, we first introduce the speech materials, processing conditions, and analysis methods used in these experiments. 

\subsubsection{Test stimuli and processing}

For the subjective listening tests, we selected 90 speech samples from the out-of-domain test set: 30 clean samples, 30 samples mixed with street noise (nonstationary environmental noise), and 30 samples containing interfering speech. The noisy samples were generated at SNRs of \(-1\) and \(5\) dB.

In the listening tests, we compared the Conventional, NeuroAMP, and AV-NeuroAMP methods (see Sec.~\ref{subsec:comparison}). Although video data was used to generate the test stimuli, the visual information was not presented to the listeners during the testing process; they listened only to the processed audio. The listening tests utilized Mandarin speech, which is from out-of-domain data. For listeners with hearing loss, the audiogram data from the test ear was used to design the compensation.

\subsubsection{Listening procedure and statistical analysis}

For each stimulus, listeners completed two evaluation tasks. Speech quality was rated without a reference signal on a scale from 0 (worst) to 5 (best), with the analysis focusing on within-listener differences across methods. Speech intelligibility was measured as the proportion of Chinese characters correctly identified in each 10-character utterance, yielding scores from 0 to 1. Listeners were allowed to replay each stimulus once; therefore, each stimulus could be presented no more than twice.

The experiment used a MacBook as the playback device, and the stimuli were presented through Sennheiser IE 200 in-ear headphones. Prior to the formal listening test, participants listened to a set of sample audio recordings to adjust the volume to a safe and comfortable level. These sample audio recordings were used only for volume adjustment and familiarization with the environment and were not included in the formal assessment. The sample audio recordings included clean speech processed with NAL-R/WDRC, mixed street noise speech with a signal-to-noise ratio of -1 dB (also processed with NAL-R/WDRC), and the same noisy speech processed with AV-NeuroAMP. These examples were selected from the out-of-domain test set and did not overlap with the sentences used in the formal listening test. During the formal assessment, the order of the stimuli was randomized.

The same analysis was applied to both experiments, with the listener as the unit of analysis in Experiment~1
(Sec.~\ref{subsec:nhmethods}) and the test ear in Experiment~2
(Sec.~\ref{subsec:himethods}). For each unit, scores were first averaged over the sentences in each of the nine method~$\times$~condition cells (three processing methods: Conventional, NeuroAMP, and AV-NeuroAMP; three acoustic conditions: clean, street noise, and competing speech). The cell means were submitted to a two-way repeated-measures analysis of variance (ANOVA) with method and condition as within-subject factors; Greenhouse--Geisser correction was applied to the degrees of freedom, and effect sizes are reported as partial $\eta^{2}$. Because the method~$\times$~condition interaction was significant for both outcomes, pairwise method comparisons were carried out within each condition using two-sided paired $t$-tests, with Holm correction applied across the nine comparisons for each outcome; effect sizes are reported as Cohen's $d_z$. Results are summarized as means, standard deviations, and 95\% $t$-based confidence intervals. Statistical significance was defined as a Holm-adjusted $p < 0.05$.

\subsubsection{Experiment 1: normal-hearing listeners under simulated hearing loss}

\label{subsec:nhmethods}
\begin{figure}[ht]
\begin{center}
\includegraphics[width=\columnwidth]{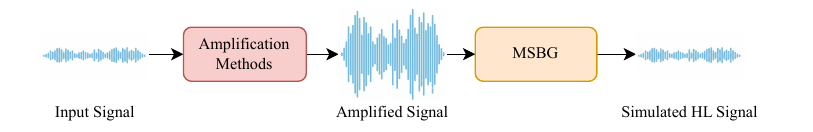}
\caption{\label{fig:msbg}{Signal path of the preliminary experiment with normal-hearing listeners: hearing-aid processing followed by MSBG simulation of the corresponding hearing loss.}}
\end{center}
\end{figure}

In the simulated hearing loss experiment, the process of generating the listening test signal is shown in Figure~\ref{fig:msbg}. We used the Moore-Stone-Baer-Glasberg (MSBG) hearing loss simulation model ~\citep{MSBG1,MSBG2}, which can generate a simulated hearing loss function based on a given audiogram. Next, we obtained a personalized amplified signal from the input speech signal based on the given audiogram, and then input the amplified signal into the hearing loss simulation function to obtain the final listening test signal. This listening test experiment was conducted by 29 subjects with normal hearing, including 14 women and 15 men, aged between 20 and 50 years. This experiment can be considered a preliminary experiment aimed at verifying the effectiveness of the proposed method compared with other methods under simulated hearing loss conditions.

\subsubsection{Experiment 2: listeners with hearing loss}
\label{subsec:himethods}
In our listening testing experiment with hearing aid users, we recruited 12 participants with hearing loss, including 8 women and 4 men, with a mean age of 65.75 years (standard deviation 12.88, age range 44–81 years). A total of 20 ears were tested: 8 participants underwent binaural testing, and 4 participants underwent unilateral testing. The test was conducted using a single ear, with each ear receiving a different test corpus. To avoid the same participant hearing the same data repeatedly, we divided a corpus of 90 sentences into two non-overlapping sets of 45 sentences each, assigned to each participant's two ears. Each set of sentences contained 15 sentences from each of the three testing methods and was processed using audiograms from the tested ear. There was a 5–10 second interval between sentences to allow participants time to respond.

Across the 20 test ears, the mean four-frequency pure-tone average(PTA4; 0.5, 1, 2, and 4~kHz)~\citep{chadha2021world} was 59.2~dB~HL (standard deviation 14.6; range 41.3--91.3~dB~HL). Mean hearing thresholds increased with frequency, from 40.2~dB~HL at 250~Hz to 56.8~dB~HL at 1~kHz, 68.8~dB~HL at 4~kHz, and 81.8~dB~HL at 8~kHz. Based on the WHO PTA4 classification~\citep{chadha2021world}, six ears had moderate hearing loss, seven had moderately severe hearing loss, five had severe hearing loss, and two had profound hearing loss. Based on audiogram configuration, 11 ears exhibited a sloping pattern, defined as a mean threshold at 2--4~kHz at least 15~dB higher than that at 250--500~Hz, whereas the remaining nine ears exhibited a flat pattern. Among the eight listeners tested bilaterally, the mean absolute interaural difference in PTA4 was 3.4~dB and did not exceed 8.8~dB, indicating broadly similar hearing-loss severity between the two ears.

\section{Results}
\label{sec:results}

This section presents the objective evaluation results, followed by the results of the two listening experiments.

\subsection{Objective evaluation results}
\label{subsec:objective}

The objective evaluation results include five parts: in-domain performance, gain error in specific frequency bands, comparison of end-to-end processing and two-stage processing, out-of-domain performance, and ablation experiment for the audiogram conditioning methods.

\subsubsection{In-domain performance}
\label{subsec:indomain}

Table ~\ref{tab:indomain} reports the in-domain objective results for all systems under clean and noisy input conditions.

Under clean input conditions, the performance of the two end-to-end systems (NeuroAMP and AV-NeuroAMP) is close to that of Conventional($\mathbf{S}_{\mathrm{Amp}}$). AV-NeuroAMP scores 0.925 and 0.932 for HASPI and HASQI, respectively; NeuroAMP scores 0.917 and 0.902, and Conventional($\mathbf{S}_{\mathrm{Amp}}$) scores 0.928 and 0.944, respectively. Furthermore, compared to NeuroAMP, AV-NeuroAMP reduced MSE from $1.6\times10^{-3}$ to $2.6\times10^{-4}$, MBGE from 1.30 dB to 0.79 dB, and increased the target reference SNR from 18.59 dB to 20.73 dB.

In noisy-input conditioning, Conventional($\mathbf{Y}_{\mathrm{Amp}}$) scored 0.449 and 0.283 on HASPI and HASQI, respectively, while NeuroAMP scored 0.419 and 0.268. AV-NeuroAMP improved these scores to 0.780 and 0.625, respectively. Compared to NeuroAMP, AV-NeuroAMP reduces MSE from $2.6\times10^{-2}$ to $1.1\times10^{-3}$, MBGE from 11.30 dB to 3.48 dB, and increases the target reference SNR from -2.98 dB to 10.20 dB.

\begin{table*}[ht]
\centering
\caption{Results of an objective evaluation of the in-domain test set. System definitions are given in Section ~\ref{subsec:comparison}, and metric definitions are given in Section ~\ref{sec:objective_metrics}; Conventional($\mathbf{S}_{\mathrm{Amp}}$) is the reference system.  Best performance under each condition, except for the reference system is shown in \textbf{bold}.}
\label{tab:indomain}
\begin{tabular}{lccccc}
\hline
\textbf{Method}
& \textbf{HASPI} $\uparrow$
& \textbf{HASQI} $\uparrow$
& \textbf{MSE} $\downarrow$
& \textbf{MBGE (dB)} $\downarrow$
& \textbf{SNR (dB)} $\uparrow$ \\
\hline
\multicolumn{6}{c}{\textit{Clean input}} \\
\hline
Conventional ($\mathbf{S}_{\mathrm{Amp}}$)
& $0.928 \pm 0.16 $ & $0.944 \pm 0.08$ & -- & -- & -- \\
Two-stage(A)
& $0.901 \pm 0.16 $ & $0.843 \pm 0.12 $ & $\mathbf{7.8\times10^{-5}}$ & $3.58 \pm 2.38 $ & $\mathbf{23.81 \pm 5.78} $ \\
Two-stage(AV)
& $0.879 \pm 0.18 $ & $0.696 \pm 0.16 $ & $9.9\times10^{-3}$ & $6.74 \pm 3.04 $ & $-0.37 \pm 5.09 $ \\
NeuroAMP
& $0.917 \pm 0.16 $ & $0.902 \pm 0.13 $ & $1.6\times10^{-3}$ & $1.30 \pm 1.93 $ & $18.59 \pm 4.94 $ \\
AV-NeuroAMP
& $\mathbf{0.925 \pm 0.15} $ & $\mathbf{0.932 \pm 0.07} $ & $2.6\times10^{-4}$ & $\mathbf{0.79 \pm 0.49} $ & $20.73 \pm 5.16 $ \\
\hline
\multicolumn{6}{c}{\textit{Noisy input}} \\
\hline
Conventional ($\mathbf{Y}_{\mathrm{Amp}}$)
& $0.449 \pm 0.34 $ & $0.283 \pm 0.20 $ & $1.9\times10^{-2}$ & $10.51 \pm 4.72 $ & $-2.04 \pm 6.57 $ \\
Two-stage(A)
& $0.482 \pm 0.38 $ & $0.363 \pm 0.28 $ & $1.0\times10^{-2}$ & $9.90 \pm 4.51 $ & $4.70 \pm 8.91 $ \\
Two-stage(AV)
& $0.590 \pm 0.31 $ & $0.317 \pm 0.20 $ & $4.6\times10^{-2}$ & $11.36 \pm 3.65 $ & $-7.66 \pm 5.20 $ \\
NeuroAMP
& $0.419 \pm 0.35 $ & $0.268 \pm 0.20 $ & $2.6\times10^{-2}$ & $11.30 \pm 5.23 $ & $-2.98 \pm 7.18 $ \\
AV-NeuroAMP
& $\mathbf{0.780 \pm 0.24} $ & $\mathbf{0.625 \pm 0.15} $ & $\mathbf{1.1\times10^{-3}}$ & $\mathbf{3.48 \pm 1.27} $ & $\mathbf{10.20 \pm 3.73} $ \\
\hline
\end{tabular}
\end{table*}

\subsubsection{Band-specific gain error}
\label{subsec:gainresults}

\begin{figure}[!ht]
\begin{center}
\includegraphics[width=\columnwidth]{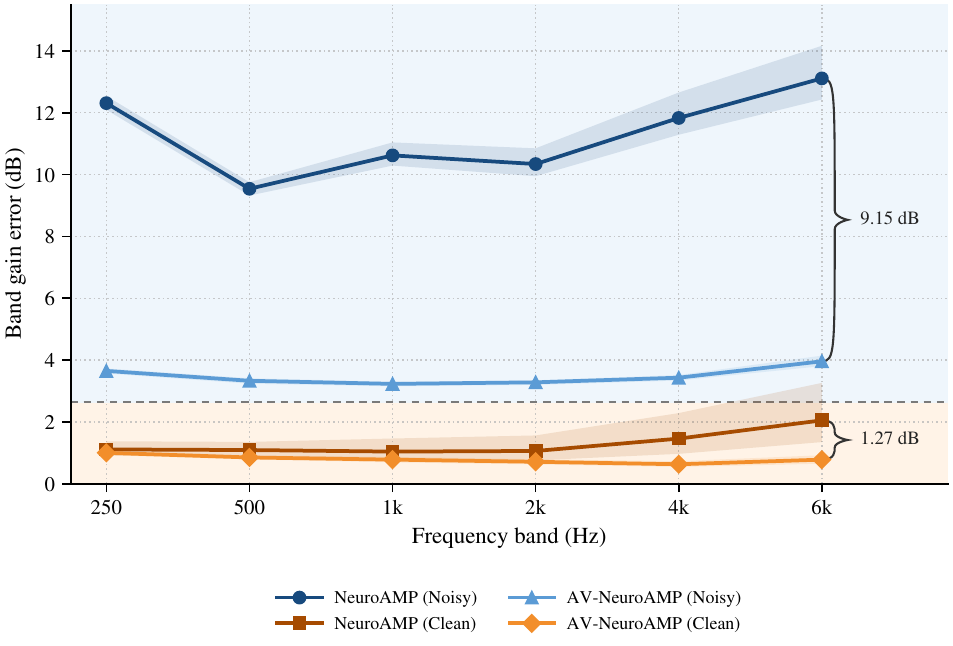}
\caption{\label{fig:gain}{Band-specific gain error relative to the listener-specific target on the in-domain test set. Blue and orange curves represent noisy- and clean-input conditions, respectively; shades and marker shapes distinguish NeuroAMP from AV-NeuroAMP, as indicated in the legend. Shaded bands around the curves indicate 95\% confidence intervals. Light blue and orange backgrounds, separated by a dashed line, visually distinguish the two input conditions. Braces mark the largest between-system difference for each condition.}}
\end{center}
\end{figure}

To examine how the MBGE in Table ~\ref{tab:indomain} varies with frequency, we summarized the band errors from the in-domain tests. Figure~\ref{fig:gain} shows the results for clean and noisy inputs.

For clean input, NeuroAMP's errors are relatively consistent across the 250–2000 Hz band (1.11, 1.09, 1.04, and 1.06 dB, respectively), while the errors are larger at 4 kHz and 6 kHz (1.46 and 2.05 dB, respectively). Under noisy conditions, the errors increase significantly and exhibit an approximately U-shaped distribution across the bands: 12.32 dB at 250 Hz, a minimum of 9.54 dB at 500 Hz, and a maximum of 13.12 dB at 6 kHz.

In contrast, the AV-NeuroAMP's errors are all below 1 dB, with a maximum error at 250 Hz and a minimum error at 6 kHz decreasing with frequency for clean input. The overall error values across all frequency bands exhibit a flat distribution, similar to the performance of NeuroAMP. Under noisy input, the AV-NeuroAMP shows a maximum error of 4 dB at 6 kHz and a minimum of 3.4 dB at 1 kHz. The overall error values across all frequency bands also show a flat distribution. Compared to NeuroAMP, AV-NeuroAMP reduces the error by 6.21-–9.16 dB in each frequency band.

\subsubsection{End-to-end versus two-stage processing}
\label{subsec:twostageresults}

Table ~\ref{tab:indomain} compares the performance of end-to-end methods (NeuroAMP and AV-NeuroAMP) with two-stage methods (Two-stage(A) and Two-stage (AV)).

For clean input, Two-stage(A) achieved the lowest MSE (\(7.8\times10^{-5}\)) and the highest target-referenced SNR (23.81~dB) among all learning-based systems; its HASPI and HASQI scores, respectively, were 0.901 and 0.843, outperforming Two-stage (AV) but remaining lower than those of NeuroAMP and AV-NeuroAMP. In contrast, Two-stage (AV) yielded the lowest HASQI score (0.696) and a negative target-referenced SNR (\(-0.37\)~dB).

For noisy input, HASPI and HASQI were 0.482 and 0.363 for the Two-stage(A) and 0.590 and 0.317 for the Two-stage (AV). AV-NeuroAMP obtained the best score on all five metrics in this condition.

\subsubsection{Out-of-domain performance}
\label{subsec:outdomainresults}

Table~\ref{tab:outdomain} presents the objective evaluation results on the Mandarin out-of-domain test set.

For clean input, the end-to-end systems, NeuroAMP and AV-NeuroAMP, achieved similar performance, with both yielding results close to those of Conventional. More specifically compared with NeuroAMP, AV-NeuroAMP achieved a lower HASPI score (0.879), a higher HASQI score (0.887), a higher MSE (\(1.4\times10^{-3}\)), a lower MBGE (1.48~dB), and a higher target-referenced SNR (19.28~dB).

For noisy input, AV-NeuroAMP achieved the best performance across all evaluation metrics. Its HASPI and HASQI scores were 0.598 and 0.454, respectively, compared with 0.475 and 0.292 for Conventional and 0.449 and 0.303 for NeuroAMP. AV-NeuroAMP also achieved an MSE of \(3.4\times10^{-3}\), an MBGE of 6.04~dB, and a target-referenced SNR of 7.35~dB, outperforming both NeuroAMP and the Conventional method.

Although AV-NeuroAMP achieved lower scores for out-of-domain noisy inputs than for in-domain noisy inputs, it still outperformed all other systems on the out-of-domain test set.

\begin{table*}[t]
\centering
\caption{Objective results on the Mandarin out-of-domain test set. Systems, metrics, and conventions are as in Table~\ref{tab:indomain}.}
\label{tab:outdomain}
\small
\setlength{\tabcolsep}{3.5pt}
\begin{tabular}{lccccc}
\hline
\textbf{Method}
& \textbf{HASPI} $\uparrow$
& \textbf{HASQI} $\uparrow$
& \textbf{MSE} $\downarrow$
& \textbf{MBGE (dB)} $\downarrow$
& \textbf{SNR (dB)} $\uparrow$ \\
\hline
\multicolumn{6}{c}{\textit{Clean input}} \\
\hline
Conventional ($\mathbf{S}_{\mathrm{Amp}}$)
& $0.896 \pm 0.21$ & $0.905 \pm 0.11$ & --  & -- & -- \\
NeuroAMP
& $\mathbf{0.884} \pm 0.22 $ & $0.877 \pm 0.13 $ & $\mathbf{8.4\times10^{-4}}$ & $1.68 \pm 1.77 $ & $16.20 \pm 4.77 $ \\
AV-NeuroAMP
& $0.879 \pm 0.23 $ & $\mathbf{0.887} \pm 0.16 $ & $1.4\times10^{-3}$ & $\mathbf{1.48} \pm 2.43 $ & $\mathbf{19.28} \pm 7.90$ \\
\hline
\multicolumn{6}{c}{\textit{Noisy input}} \\
\hline
Conventional ($\mathbf{Y}_{\mathrm{Amp}}$)
& $0.475 \pm 0.31$ & $0.292 \pm 0.15 $ & $9.5\times10^{-3}$ & $10.84 \pm 3.29 $ & $3.08 \pm 4.55 $ \\
NeuroAMP
& $0.449 \pm 0.31 $ & $0.303 \pm 0.15 $ & $6.0\times10^{-3}$ & $9.36 \pm 2.40 $ & $4.21 \pm 3.05 $ \\
AV-NeuroAMP
& $\mathbf{0.598 \pm 0.31} $ & $\mathbf{0.454 \pm 0.18} $ & $\mathbf{3.4\times10^{-3}}$ & $\mathbf{6.04 \pm 2.00} $ & $\mathbf{7.35 \pm 3.12} $ \\
\hline
\end{tabular}
\end{table*}

\subsubsection{AC-FiLM versus direct concatenation}
\label{subsec:ablationresults}

Table~\ref{tab:ablation} shows the ablation experimental results for the two audiogram conditioning methods. As shown in the table, regardless of whether the input signal is clean or noisy, AC-FiLM outperforms direct concatenation in all five metrics. With a clean input signal, compared to direct concatenation, AC-FiLM improves HASPI from 0.920 to 0.925, HASQI from 0.913 to 0.932, MSE from $3.7\times10^{-4}$ to $2.6\times10^{-4}$, MBGE from 1.07 dB to 0.79 dB, and the target reference signal-to-noise ratio (SNR) from 19.01 dB to 20.73 dB. For noisy input, HASPI increased from 0.738 to 0.780, and HASQI increased from 0.563 to 0.625. MSE decreased from $1.4\times10^{-3}$ to $1.1\times10^{-3}$, MBGE decreased from 4.15 dB to 3.48 dB, while the target reference SNR increased from 8.85 dB to 10.20 dB.

\begin{figure}[ht]
\begin{center}
\includegraphics[width=\columnwidth]{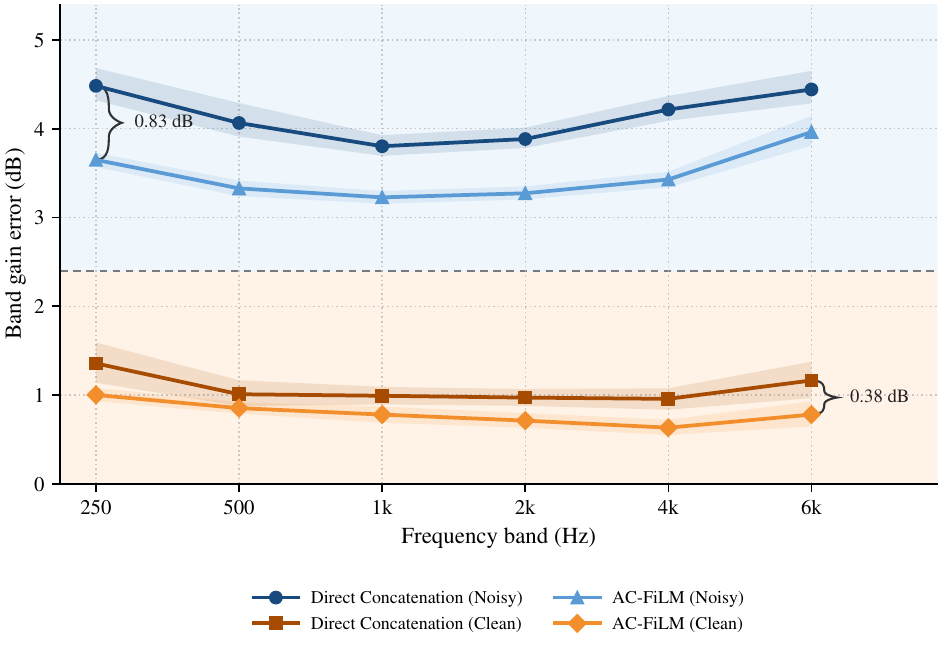}
\caption{\label{fig:ablgain}
Band-specific gain error for the two audiogram-conditioning strategies, over the in-domain test set.  Blue and orange curves represent noisy- and clean-input conditions, respectively; shades and marker shapes distinguish the two conditioning strategies. Shaded bands around the curves indicate 95\% confidence intervals. Light blue and orange backgrounds, separated by a dashed line, visually distinguish the two input conditions. Braces mark the largest between-strategy difference for each condition.}
\end{center}
\end{figure}

Figure~\ref{fig:ablgain} shows the band-specific results of the AC-FiLM exhibits lower errors across all bands under both clean and noisy input conditions. Under clean input conditions, the AC-FiLM effectively reduces gain error by 0.16 dB to 0.38 dB compared to direct concatenation, from 500 Hz to 6 kHz. Under noisy input conditions, the gain error is reduced by 0.48 dB to 0.83 dB (250 Hz to 6 kHz).

\begin{table*}[t]
\centering
\renewcommand{\arraystretch}{1.15}
\caption{Audiogram-conditioning ablation study on the in-domain test set, using the same audiovisual backbone and training and evaluation protocols (Sec.~\ref{subsec:ablationdesign}). Reporting conventions follow Table~\ref{tab:indomain}. The best score for each metric is shown in \textbf{bold}.}
\label{tab:ablation}
\small
\setlength{\tabcolsep}{3.5pt}
\begin{tabular}{lccccc}
\hline
\textbf{Fusion strategy}
& \textbf{HASPI} $\uparrow$
& \textbf{HASQI} $\uparrow$
& \textbf{MSE} $\downarrow$
& \textbf{MBGE (dB)} $\downarrow$
& \textbf{SNR (dB)} $\uparrow$ \\
\hline
\multicolumn{6}{c}{\textit{Clean input}} \\
\hline
Direct Concatenation
& $0.920 \pm 0.16 $ & $0.913 \pm 0.08 $ & $3.7\times10^{-4}$ & $1.07 \pm 0.83 $ & $19.01 \pm 5.43 $ \\
AC-FiLM
& $\mathbf{0.925} \pm 0.15 $ & $\mathbf{0.932} \pm 0.07 $ & $\mathbf{2.6\times10^{-4}}$ & $\mathbf{0.79} \pm 0.49 $ & $\mathbf{20.73} \pm 5.16 $ \\
\hline
\multicolumn{6}{c}{\textit{Noisy input}} \\
\hline
Direct Concatenation
& $0.738 \pm 0.25 $ & $0.563 \pm 0.16 $ & $1.4\times10^{-3}$ & $4.15 \pm 1.43 $ & $8.85 \pm 3.68 $ \\
AC-FiLM
& $\mathbf{0.780} \pm 0.24 $ & $\mathbf{0.625} \pm 0.15 $ & $\mathbf{1.1\times10^{-3}}$ & $\mathbf{3.48} \pm 1.27 $ & $\mathbf{10.20} \pm 3.73 $ \\
\hline
\end{tabular}
\end{table*}

\subsection{Listening-test results}
\label{subsec:listeningresults}

In this section, we first present the listening-test results for normal-hearing participants under simulated hearing loss, followed by those for hearing-aid users.

\subsubsection{Experiment 1: normal-hearing listeners under simulated hearing loss}
\label{subsec:nhresults}

\begin{figure}[ht] 
\begin{center}
\figcolumn{
\fig{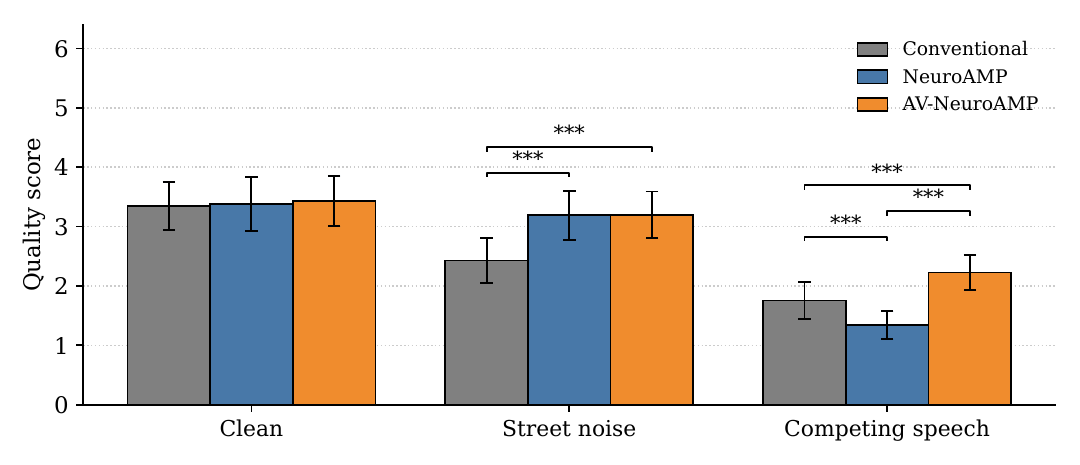}{.5\textwidth}{(A)\label{fig:FIG7a}}
\fig{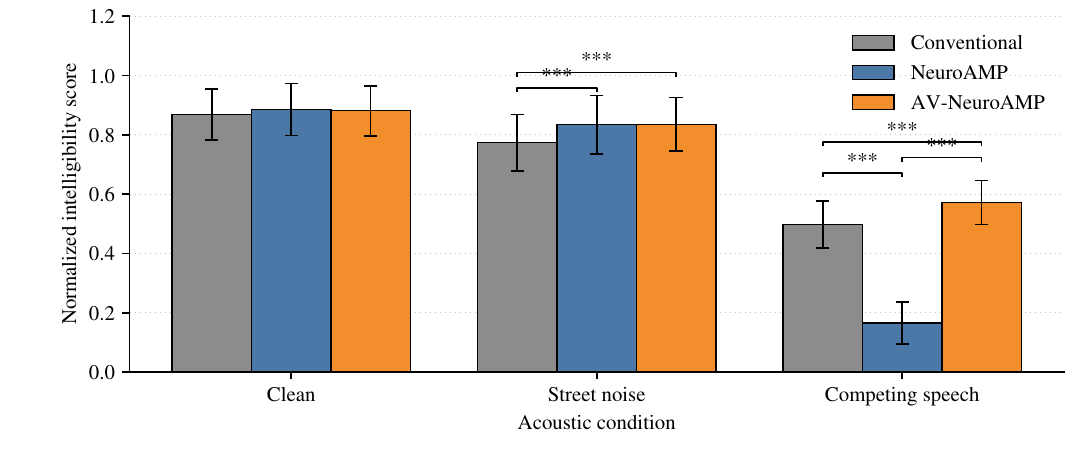}{.5\textwidth}{(B)\label{fig:FIG7b}}
}
\caption{\label{fig:nh}{(A) Speech quality scores, top, and (B) speech intelligibility scores, bottom, for the preliminary experiment with normal-hearing listeners under MSBG-simulated hearing loss ($n=29$). Quality was rated on a 0--5 scale; the intelligibility score is the proportion of correctly identified Mandarin characters, ranging from 0 to 1. Bars are means and error bars are 95\% confidence intervals. Holm correction was applied across the pairwise comparisons within each outcome, and only significant pairs are annotated: $^{*}p<0.05$, $^{**}p<0.01$, $^{***}p<0.001$.}}
\end{center}
\end{figure}

Figure~\ref{fig:nh} shows the results of MSBG simulated hearing loss in 29 normal-hearing subjects.

In clean speech, we observed no significant differences among the three methods (Conventional, NeuroAMP, and AV-NeuroAMP): quality scores ranged from 3.35 to 3.43, and intelligibility scores ranged from 0.868 to 0.885.

Under street-noise conditions, both end-to-end systems outperformed Conventional in terms of speech quality and intelligibility, with significance tests confirming that all improvements were statistically significant (\(p<0.001\)). No statistically significant differences were observed between NeuroAMP and AV-NeuroAMP for either metric. Specifically, the quality scores for the Conventional method, NeuroAMP, and AV-NeuroAMP were \(2.43\pm0.98\), \(3.19\pm1.09\), and \(3.20\pm1.04\), respectively, while the corresponding intelligibility scores were \(0.774\pm0.250\), \(0.834\pm0.259\), and \(0.836\pm0.239\), respectively.

Under competing-speech conditions, NeuroAMP performed significantly worse than Conventional, whereas AV-NeuroAMP significantly outperformed both methods in terms of speech quality and intelligibility (\(p<0.001\)). Specifically, the quality scores for Conventional, NeuroAMP, and AV-NeuroAMP were \(1.75\pm0.81\), \(1.34\pm0.62\), and \(2.23\pm0.78\), respectively, while the corresponding intelligibility scores were \(0.497\pm0.210\), \(0.165\pm0.188\), and \(0.572\pm0.195\), respectively.

\subsubsection{Experiment 2: listeners with hearing loss}
\label{subsec:hiresults}

\begin{figure}[ht]
\figcolumn{
\fig{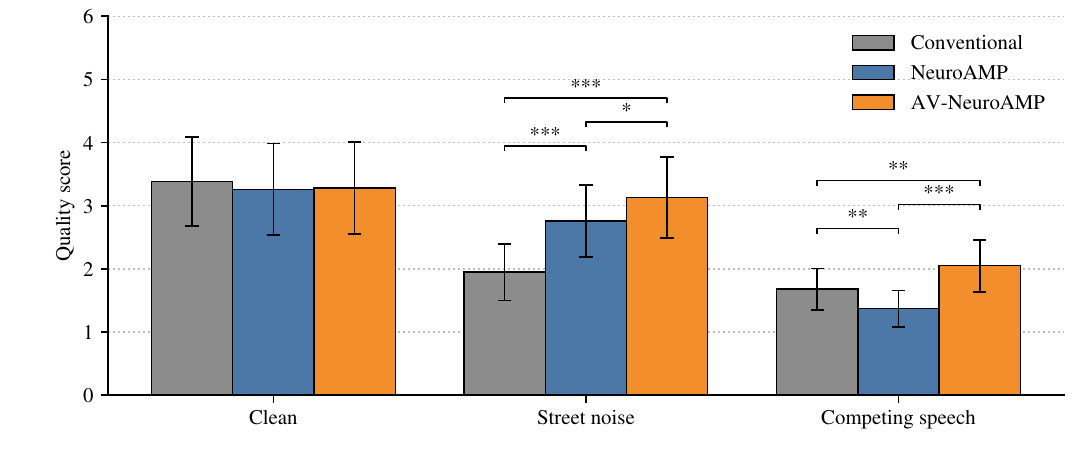}{.5\textwidth}{(A)\label{fig:FIG8a}}
\fig{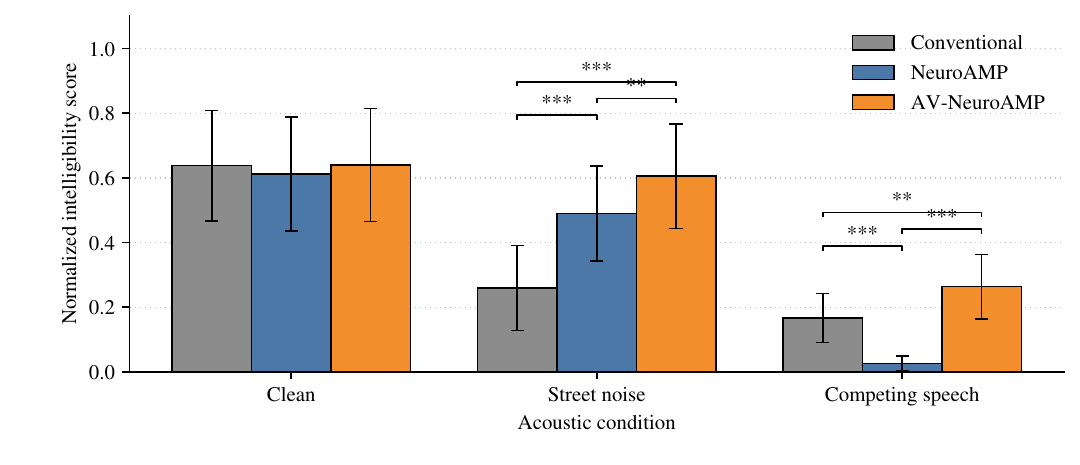}{.5\textwidth}{(B)\label{fig:FIG8b}}
}
\caption{\label{fig:hi}
Subjective listening-test results for hearing-aid users (12 listeners, 20 test ears): (A) speech-quality ratings and (B) speech-intelligibility scores. Quality was rated on a 0--5 scale, whereas intelligibility was measured as the proportion of correctly identified Mandarin characters, ranging from 0 to 1. Bars represent means, and error bars indicate 95\% confidence intervals. Significant improvements are annotated: $^{*}p<0.05$, $^{**}p<0.01$, and $^{***}p<0.001$.}
\end{figure}

Figure~\ref{fig:hi} shows the subjective listening test results of 12 patients with hearing loss (20 test ears in total).

In clean speech, we observed no significant differences among the three methods (Conventional, NeuroAMP, and AV-NeuroAMP) (all Holm-adjusted $p>0.05$). Quality scores were $3.38\pm1.51$, $3.26\pm1.55$, and $3.28\pm1.55$, respectively, while intelligibility scores were \(0.638\pm0.365\), \(0.612\pm0.378\), and \(0.640\pm0.374\), respectively.

Under street-noise conditions, the quality scores for Conventional, NeuroAMP, and AV-NeuroAMP were \(1.95\pm0.96\), \(2.76\pm1.21\), and \(3.13\pm1.36\), respectively. Compared with Conventional, NeuroAMP and AV-NeuroAMP improved the quality score by 0.81 and 1.18 points, respectively. AV-NeuroAMP also achieved a 0.37-point improvement over NeuroAMP. All three pairwise differences in quality were statistically significant after Holm correction \((p_{\mathrm{Holm}}<0.05)\). The corresponding normalized intelligibility scores were \(0.260\pm0.281\), \(0.490\pm0.314\), and \(0.606\pm0.346\), respectively. Relative to Conventional, NeuroAMP and AV-NeuroAMP improved the normalized intelligibility score by 0.230 and 0.346, respectively, while AV-NeuroAMP improved it by 0.116 relative to NeuroAMP. All three pairwise differences in intelligibility were statistically significant after Holm correction \((p_{\mathrm{Holm}}<0.01)\).

In competing speech, AV-NeuroAMP achieved the highest quality and intelligibility scores, while NeuroAMP achieved the lowest. The quality scores for Conventional, NeuroAMP, and AV-NeuroAMP were $1.68\pm0.70$, $1.37\pm0.62$, and $2.05\pm0.88$, respectively. while the intelligibility scores were $0.167\pm0.161$, $0.027\pm0.050$, and $0.264\pm0.212$, respectively. Compared to NeuroAMP, AV-NeuroAMP improved quality and intelligibility by 0.68 and 0.237 points, respectively. Compared to Conventional, AV-NeuroAMP improved quality and intelligibility by 0.37 and 0.097 points, respectively. Conventional also outperformed NeuroAMP by 0.31 points in quality and 0.140 points in intelligibility. All pairwise differences were statistically significant after Holm correction (all $p_{\mathrm{Holm}}<0.01$).

\section{Discussion}
\label{sec:discussion}

This study investigated whether an end-to-end audiovisual neural amplifier could jointly perform SE and listener-specific hearing-loss compensation, particularly under competing-speech conditions, where audio-only processing remains challenging. The results support three main conclusions. First, AV-NeuroAMP provided substantial objective and perceptual benefits under noisy conditions while maintaining performance close to the prescribed target for clean speech. Second, its advantage was strongly dependent on the acoustic condition and was most pronounced under competing-speech interference. Third, AC-FiLM provided more effective audiogram conditioning than direct concatenation within the same audiovisual backbone. Together, these findings suggest that combining visual cues with effective audiogram conditioning enables end-to-end neural amplification to better address both target-speech enhancement and listener-specific hearing-loss compensation, with the greatest perceptual benefits observed under competing-speech conditions.

\subsection{Perceptual benefits given by visual cues}
\label{subsec:conditiondisc}
\label{subsec:targetselection}

In the subjective listening tests, hearing-aid users exhibited trends consistent with those observed in normal-hearing participants under simulated hearing loss. Under clean-speech conditions, no significant differences were observed among the three methods in either experiment. These findings indicate that, in quiet environments, neither end-to-end learning-based method degrades speech quality or intelligibility. Thus, both methods effectively provide listener-specific amplification while achieving performance similar to that of Conventional.

Under street-noise conditions, both end-to-end learning-based methods significantly improved speech quality and intelligibility compared with Conventional. In the listening tests involving hearing-aid users, AV-NeuroAMP significantly outperformed NeuroAMP on both metrics. These results indicate that audio-only end-to-end processing can improve the listening experience in background noise, while the proposed AV-NeuroAMP, incorporating AC-FiLM for more effective audiogram conditioning, can provide further improvements in speech quality and intelligibility.

The differences among the three methods were most pronounced under competing-speech conditions. In both listening tests, AV-NeuroAMP achieved the highest speech-quality and intelligibility scores, followed by Conventional and then NeuroAMP, with all pairwise differences being statistically significant. NeuroAMP’s poorer performance relative to Conventional indicates that audio-only end-to-end neural amplification is not inherently advantageous in the presence of competing speech. Because the target and interfering speech share similar spectral and temporal characteristics, an audio-only model may inadvertently preserve the interfering speech or suppress portions of the target speech, thereby reducing target-speech intelligibility. Although Conventional cannot separate the target speech from the interference, it avoids the target-selection errors and associated distortions that may arise from imperfect enhancement. In contrast, AV-NeuroAMP uses facial movements as complementary articulatory cues to identify and extract the target speech more accurately, thereby reducing interference amplification and target-speech distortion. Notably, listeners heard only the processed audio during the experiments; therefore, the observed benefits did not result from audiovisual integration by the listeners themselves. Rather, AV-NeuroAMP used visual information to guide signal generation and identify the target speaker. These findings demonstrate that visual information can facilitate hearing-aid signal processing even when no visual content is presented to the user.

\subsection{End-to-end processing and out-of-domain performance}
\label{subsec:jointdisc}
\label{subsec:crossdomain}

The in-domain objective results support the direct optimization of the final listener-specific output under noisy conditions. AV-NeuroAMP outperformed all evaluated audio-only and audiovisual two-stage systems across the five objective metrics. One likely advantage of end-to-end processing is that the model is optimized directly toward the final NAL-R/WDRC target. In contrast, the SE component of a two-stage system is optimized to estimate clean speech without explicitly accounting for subsequent listener-specific amplification and compression. Consequently, residual interference, speech distortion, or level errors introduced during the SE stage may propagate to or be amplified by the compensation stage. Joint optimization enables SE, amplification, and compression to be coordinated within a unified mapping.

The results indicate that audio-only end-to-end processing is not superior under every condition or evaluation metric. For clean input, the audio-only two-stage system achieved the lowest MSE and highest target-referenced SNR among all systems, whereas AV-NeuroAMP yielded higher HASPI and HASQI scores and a lower MBGE. These differences illustrate that small signal-level errors do not necessarily correspond to better predicted perceptual quality or more accurate short-time gain across frequency bands. AV-NeuroAMP maintained its advantage on the out-of-domain Mandarin dataset, achieving the best noisy-input performance across all five objective metrics. This finding is particularly encouraging because the training and evaluation datasets differed in language, speakers, and acoustic conditions.

\subsection{Audiogram conditioning and gain accuracy}
\label{subsec:acfilmdisc}

The audiogram-conditioning ablation study provides direct evidence for the effectiveness of AC-FiLM relative to direct concatenation. Because both systems used the same audiovisual backbone, training data, target, loss function, and evaluation protocol, the improvements can be attributed to the conditioning strategy as a whole. AC-FiLM outperformed direct concatenation on all five metrics for both clean and noisy input. Its lower MBGE further indicates that the model output more closely followed the listener-specific target levels over short-time frequency bands. The band-specific analysis confirmed lower gain errors across all evaluated bands, rather than an improvement confined to only one audiometric region.

These findings are consistent with the motivation for applying audiogram conditioning after instance normalization. Direct concatenation treats the audiogram as an additional input feature and requires subsequent layers to learn its interaction with the audiovisual representation implicitly. Absolute offsets or scales associated with the audiogram may be attenuated when the fused representation is subsequently normalized. AC-FiLM employs local and global conditional branches to transform the normalized audiogram into audiogram frequency details and global conditional features, which are then processed by a control module to obtain scaling, shifting, and gating parameters. The gating residual path allows the model to incorporate listener-specific modifications without discarding crucial information from the original audiovisual representation.

\subsection{Limitations and future work}
\label{subsec:limitations}
\label{subsec:broader}

Several limitations are noted and should be investigated in future work. First, while the perceptual results support the full scope of the AV-NeuroAMP system, the current framework assumes that the provided video corresponds to the intended speaker. In real-world multi-person conversational settings, goals may need to be inferred from gaze, head orientation, user choice, or other behavioral cues~\citep{yang2026tracking}. Expanding the framework to incorporate these cues could allow hearing aids to adapt not only to the listener's hearing loss but also to the listener's current communication goals, consistent with the broader goals of personalized artificial hearing systems~\citep{lesica2021ai}.

Second, consistency with the NAL-R/WDRC goal does not imply that the goal is optimal for all listeners. NAL-R is a linear prescription, while current fittings typically use non-linear prescriptions, such as NAL-NL2~\citep{Nal-nl2}. Therefore, the current results should be interpreted as evidence that AV-NeuroAMP can learn selected listener-specific processing goals, rather than evidence of optimal clinical fitting. Future research should evaluate non-linear prescriptions, personalized comfort constraints, alternative compression settings, and direct optimization based on perception or intelligibility. Monaural evaluation also does not address binaural processing, including the preservation of binaural cues that may be affected by nonlinear processing~\citep{brown2016binaural,peng2022spatialrelease}.

Finally, the current system processes complete utterances and is not yet designed for causal wearable operation. Practical deployment will require a streaming implementation with controlled algorithmic latency, memory use, and power consumption. The visual encoder accounts for most of the parameters and about 41\% of the computational cost, suggesting that lightweight visual processing, model compression, quantization, pruning, distillation, or event-triggered visual analysis may be needed~\citep{martel2023lightweight}. Given recent advances in AI algorithms~\citep{chao2026rtsemamba} and hardware~\citep{ni2026fpga}, we believe that the proposed AV-NeuroAMP system could be implemented on wearable devices, such as smart glasses or earphones, to enable real-time processing in the near future.

\section{Conclusions}
\label{sec:conclusion}

AV-NeuroAMP extends personalized neural amplification with target-talker video and audiogram-conditioning feature modulation. This Mamba-based end-to-end model uses noisy audio, synchronized video, and an audiogram to estimate the final listener-specific NAL-R/WDRC target, jointly learning SE and hearing-loss compensation.

The main findings were:

\begin{itemize}
\item With noisy input, AV-NeuroAMP outperformed Conventional and audio-only NeuroAMP on all five objective metrics in both English in-domain and Mandarin out-of-domain evaluations.
\item Under noisy in-domain conditions, AV-NeuroAMP outperformed the two-stage SE--amplification cascades on all five objective metrics, supporting direct estimation of the final listener-specific target.
\item AC-FiLM outperformed direct audiogram concatenation on all five objective metrics with in-domain clean and noisy input, including closer agreement with the listener-specific multiband gain target.
\item In listeners with hearing loss, the method-by-condition interaction was significant. No significant method differences were detected for clean speech. In street noise, all three pairwise method comparisons were significant for both quality and intelligibility, with AV-NeuroAMP achieving the highest scores, followed by NeuroAMP and Conventional. In competing speech, AV-NeuroAMP achieved the highest quality and intelligibility scores, followed by Conventional and NeuroAMP, with all pairwise differences reaching significance.
\end{itemize}

Collectively, these results support extending existing neural amplification frameworks with visual cues to better address challenging listening conditions, particularly competing speech, relative to the evaluated NeuroAMP baseline.

Future work should focus on developing causal, low-latency implementations for wearable devices, evaluating the framework through larger counterbalanced speech-recognition trials, and exploring its applications to cochlear implants and closed-loop neural or cognitive conditioning.

\section*{AUTHOR DECLARATIONS}

\subsection*{Conflict of Interest}
F.G.Z. owns stock in Axonics, Neocortix, Nurotron, Syntiant, and Xense. The other authors declare no competing interests. All co-authors have seen and agree with the contents of the manuscript. We certify that the submission is original work and is not under review at any other publication.

\subsection*{Ethics Approval}
All experimental procedures involving human participants were approved by the Institutional Review Board on Biomedical Science Research, Academia Sinica (protocol No.\ AS-IRB-BM-25049). All participants provided written informed consent prior to participation, and the experiments were conducted in accordance with the approved protocol.

\section*{DATA AVAILABILITY}
\label{sec:DATA AVAILABILITY}
The audiovisual corpora used in this study are the Taiwan Mandarin Speech with Video corpus~\citep{TMSV} and the second Audio-Visual Speech Enhancement Challenge dataset~\citep{10023284}. Audiograms used for training and objective evaluation were obtained from the Clarity Challenge~\citep{graetzer2021clarity}.

\bibliography{references_JASA}

@article{ulyanov2016instance,
  title={Instance normalization: The missing ingredient for fast stylization},
  author={D. Ulyanov and A. Vedaldi and V. Lempitsky},
  journal={arXiv preprint arXiv:1607.08022},
  year={2016}
}

@misc{WHO_hearing_loss_2024,
  author        = {{World Health Organization}},
  title         = {Deafness and hearing loss},
  year          = {2026},
  url           = {https://www.who.int/news-room/fact-sheets/detail/deafness-and-hearing-loss}
}

@article{shukla2020hearing,
  author        = {A. Shukla and M. Harper and E. Pedersen and A. Goman and J. J. Suen and C. Price and J. Applebaum and M. Hoyer and F. R. Lin and N. S. Reed},
  title         = {Hearing loss, loneliness, and social isolation: A systematic review},
  journal       = {Otolaryngol. Head Neck Surg.},
  volume        = {162},
  number        = {5},
  pages         = {622--633},
  year          = {2020},
  doi           = {10.1177/0194599820910377}
}

@article{lawrence2020hearing,
  author        = {B. J. Lawrence and D. M. P. Jayakody and R. J. Bennett and R. H. Eikelboom and N. Gasson and P. L. Friedland},
  title         = {Hearing loss and depression in older adults: A systematic review and meta-analysis},
  journal       = {Gerontologist},
  volume        = {60},
  number        = {3},
  pages         = {e137--e154},
  year          = {2020}
}

@article{ciorba2012impact,
  author        = {A. Ciorba and C. Bianchini and S. Pelucchi and A. Pastore},
  title         = {The impact of hearing loss on the quality of life of elderly adults},
  journal       = {Clin. Interv. Aging},
  volume        = {7},
  pages         = {159--163},
  year          = {2012}
}

@article{livingston2024dementia,
  author  = {G. Livingston and J. Huntley and K. Y. Liu and S. G. Costafreda and G. Selbaek and S. Alladi and D. Ames and S. Banerjee and A. Burns and C. Brayne and N. C. Fox and C. P. Ferri and L. N. Gitlin and R. Howard and H. C. Kales and M. Kivimaki and E. B. Larson and N. Nakasujja and K. Rockwood and Q. Samus and K. Shirai and A. Singh-Manoux and L. S. Schneider and S. Walsh and Y. Yao and A. Sommerlad and N. Mukadam},
  title   = {Dementia prevention, intervention, and care: 2024 report of the Lancet standing Commission},
  journal = {Lancet},
  volume  = {404},
  number  = {10452},
  pages   = {572--628},
  year    = {2024}
}

@article{lin2013hearing,
  author        = {F. R. Lin and K. Yaffe and J. Xia and Q.-L. Xue and T. B. Harris and E. Purchase-Helzner and S. Satterfield and H. N. Ayonayon and L. Ferrucci and E. M. Simonsick},
  title         = {Hearing loss and cognitive decline in older adults},
  journal       = {JAMA Intern. Med.},
  volume        = {173},
  number        = {4},
  pages         = {293--299},
  year          = {2013}
}

@book{HA2012,
  author        = {H. Dillon},
  title         = {Hearing aids},
  publisher     = {Thieme Medical Publishers},
  year          = {2012}
}

@article{jenkins2018amplification,
  author        = {K. A. Jenkins and C. Fodor and A. Presacco and S. Anderson},
  title         = {Effects of amplification on neural phase locking, amplitude, and latency to a speech syllable},
  journal       = {Ear Hear.},
  volume        = {39},
  number        = {4},
  pages         = {810--824},
  year          = {2018},
  doi           = {10.1097/AUD.0000000000000538}
}

@article{Gygi2016,
  author        = {B. Gygi and D. A. Hall},
  title         = {Background sounds and hearing-aid users: A scoping review},
  journal       = {Int. J. Audiol.},
  volume        = {55},
  number        = {1},
  pages         = {1--10},
  year          = {2016},
  doi           = {10.3109/14992027.2015.1072773}
}

@article{Hohmann2023,
  author        = {B. Edwards},
  title         = {The future of hearing aid technology},
  journal       = {Trends Amplif.},
  volume        = {11},
  number        = {1},
  pages         = {31--45},
  year          = {2007},
  doi           = {10.1177/1084713806298004}
}

@inproceedings{akeroyd20232nd,
  author        = {M. A. Akeroyd and W. Bailey and J. Barker and T. J. Cox and J. F. Culling and S. Graetzer and G. Naylor and Z. Podwi{\'n}ska and Z. Tu},
  title         = {The 2nd {Clarity} enhancement challenge for hearing aid speech intelligibility enhancement: Overview and outcomes},
  booktitle     = {Proc. ICASSP},
  pages         = {1--5},
  year          = {2023}
}

@article{McCormack2013,
  author        = {A. McCormack and H. Fortnum},
  title         = {Why do people fitted with hearing aids not wear them?},
  journal       = {Int. J. Audiol.},
  volume        = {52},
  number        = {5},
  pages         = {360--368},
  year          = {2013}
}

@article{Nal-R,
  author        = {D. Byrne and H. Dillon},
  title         = {The {National Acoustic Laboratories}' ({NAL}) new procedure for selecting the gain and frequency response of a hearing aid},
  journal       = {Ear Hear.},
  volume        = {7},
  number        = {4},
  pages         = {257--265},
  year          = {1986},
  doi           = {10.1097/00003446-198608000-00007}
}

@article{Nal-nl1,
  author        = {D. Byrne and H. Dillon and T. Ching and R. Katsch and G. Keidser},
  title         = {{NAL-NL1} procedure for fitting nonlinear hearing aids: Characteristics and comparisons with other procedures},
  journal       = {J. Am. Acad. Audiol.},
  volume        = {12},
  number        = {1},
  pages         = {37--51},
  year          = {2001}
}

@article{Nal-nl2,
  author        = {G. Keidser and H. Dillon and M. Flax and T. Ching and S. Brewer},
  title         = {The {NAL-NL2} prescription procedure},
  journal       = {Audiol. Res.},
  volume        = {1},
  number        = {1},
  pages         = {e24},
  year          = {2011},
  doi           = {10.4081/audiores.2011.e24}
}

@article{dsl,
  author        = {S. Scollie and R. Seewald and L. Cornelisse and S. Moodie and M. Bagatto and D. Laurnagaray and S. Beaulac and J. Pumford},
  title         = {The {Desired Sensation Level} multistage input/output algorithm},
  journal       = {Trends Amplif.},
  volume        = {9},
  number        = {4},
  pages         = {159--197},
  year          = {2005},
  doi           = {10.1177/108471380500900403}
}

@article{wdrc,
  author        = {L. M. Jenstad and R. C. Seewald and L. E. Cornelisse and J. Shantz},
  title         = {Comparison of linear gain and wide dynamic range compression hearing aid circuits: Aided speech perception measures},
  journal       = {Ear Hear.},
  volume        = {20},
  number        = {2},
  pages         = {117--126},
  year          = {1999},
  doi           = {10.1097/00003446-199904000-00003}
}

@article{moore2008highfrequency,
  author        = {B. C. J. Moore and M. A. Stone and C. F{\"u}llgrabe and B. R. Glasberg and S. Puria},
  title         = {Spectro-temporal characteristics of speech at high frequencies, and the potential for restoration of audibility to people with mild-to-moderate hearing loss},
  journal       = {Ear Hear.},
  volume        = {29},
  number        = {6},
  pages         = {907--922},
  year          = {2008},
  doi           = {10.1097/AUD.0b013e31818246f6}
}

@article{openMHA,
  author        = {H. Kayser and T. Herzke and P. Maanen and M. Zimmermann and G. Grimm and V. Hohmann},
  title         = {Open community platform for hearing aid algorithm research: Open master hearing aid ({openMHA})},
  journal       = {SoftwareX},
  volume        = {17},
  pages         = {100953},
  year          = {2022}
}

@article{browning2019effects,
  author        = {J. M. Browning and E. Buss and M. Flaherty and T. Vallier and L. J. Leibold},
  title         = {Effects of adaptive hearing aid directionality and noise reduction on masked speech recognition for children who are hard of hearing},
  journal       = {Am. J. Audiol.},
  volume        = {28},
  number        = {1},
  pages         = {101--113},
  year          = {2019}
}

@article{boll1979spectral,
  author        = {S. Boll},
  title         = {Suppression of acoustic noise in speech using spectral subtraction},
  journal       = {IEEE Trans. Acoust. Speech Signal Process.},
  volume        = {27},
  number        = {2},
  pages         = {113--120},
  year          = {1979},
  doi           = {10.1109/TASSP.1979.1163209}
}

@inproceedings{berouti1979enhancement,
  author        = {M. Berouti and R. Schwartz and J. Makhoul},
  title         = {Enhancement of speech corrupted by acoustic noise},
  booktitle     = {Proc. ICASSP},
  volume        = {4},
  pages         = {208--211},
  year          = {1979},
  doi           = {10.1109/ICASSP.1979.1170788}
}

@inproceedings{scalart1996wiener,
  author        = {P. Scalart and J. V. Filho},
  title         = {Speech enhancement based on a priori signal-to-noise estimation},
  booktitle     = {Proc. ICASSP},
  volume        = {2},
  pages         = {629--632},
  year          = {1996},
  doi           = {10.1109/ICASSP.1996.543199}
}

@article{mmse1,
  author        = {Y. Ephraim and D. Malah},
  title         = {Speech enhancement using a minimum mean-square error short-time spectral amplitude estimator},
  journal       = {IEEE Trans. Acoust. Speech Signal Process.},
  volume        = {32},
  number        = {6},
  pages         = {1109--1121},
  year          = {1984}
}

@article{mmse2,
  author        = {I. Y. Soon and S. N. Koh and C. K. Yeo},
  title         = {Improved noise suppression filter using self-adaptive estimator of probability of speech absence},
  journal       = {Signal Process.},
  volume        = {75},
  number        = {2},
  pages         = {151--159},
  year          = {1999}
}

@article{martin2001noisepsd,
  author        = {R. Martin},
  title         = {Noise power spectral density estimation based on optimal smoothing and minimum statistics},
  journal       = {IEEE Trans. Speech Audio Process.},
  volume        = {9},
  number        = {5},
  pages         = {504--512},
  year          = {2001},
  doi           = {10.1109/89.928915}
}

@article{cohen2001nonstationary,
  author        = {I. Cohen and B. Berdugo},
  title         = {Speech enhancement for non-stationary noise environments},
  journal       = {Signal Process.},
  volume        = {81},
  number        = {11},
  pages         = {2403--2418},
  year          = {2001}
}

@article{cherry1953cocktail,
  author        = {E. C. Cherry},
  title         = {Some experiments on the recognition of speech, with one and with two ears},
  journal       = {J. Acoust. Soc. Am.},
  volume        = {25},
  number        = {5},
  pages         = {975--979},
  year          = {1953}
}

@article{wang2018supervised,
  author        = {D. Wang and J. Chen},
  title         = {Supervised speech separation based on deep learning: An overview},
  journal       = {IEEE/ACM Trans. Audio Speech Lang. Process.},
  volume        = {26},
  number        = {10},
  pages         = {1702--1726},
  year          = {2018}
}

@article{vanveen1988beamforming,
  author        = {B. D. {Van Veen} and K. M. Buckley},
  title         = {Beamforming: A versatile approach to spatial filtering},
  journal       = {IEEE ASSP Mag.},
  volume        = {5},
  number        = {2},
  pages         = {4--24},
  year          = {1988}
}

@article{doclo2015multichannel,
  author        = {S. Doclo and W. Kellermann and S. Makino and S. E. Nordholm},
  title         = {Multichannel signal enhancement algorithms for assisted listening devices: Exploiting spatial diversity using multiple microphones},
  journal       = {IEEE Signal Process. Mag.},
  volume        = {32},
  number        = {2},
  pages         = {18--30},
  year          = {2015}
}

@article{best2017dynamic,
  author        = {V. Best and E. Roverud and T. Streeter and C. R. Mason and Kidd, Jr., G.},
  title         = {The benefit of a visually guided beamformer in a dynamic speech task},
  journal       = {Trends Hear.},
  volume        = {21},
  year          = {2017},
  doi           = {10.1177/2331216517722304}
}

@article{spriet2005robustness,
  author        = {A. Spriet and M. Moonen and J. Wouters},
  title         = {Robustness analysis of multichannel Wiener filtering and generalized sidelobe cancellation for multimicrophone noise reduction in hearing aid applications},
  journal       = {IEEE Trans. Speech Audio Process.},
  volume        = {13},
  number        = {4},
  pages         = {487--503},
  year          = {2005}
}

@article{zakis2011wind,
  author        = {J. A. Zakis},
  title         = {Wind noise at microphones within and across hearing aids at wind speeds below and above microphone saturation},
  journal       = {J. Acoust. Soc. Am.},
  volume        = {129},
  number        = {6},
  pages         = {3897--3907},
  year          = {2011}
}

@article{hinton2012deep,
  author        = {G. Hinton and L. Deng and D. Yu and G. E. Dahl and A.-R. Mohamed and N. Jaitly and A. Senior and V. Vanhoucke and P. Nguyen and T. N. Sainath and B. Kingsbury},
  title         = {Deep neural networks for acoustic modeling in speech recognition: The shared views of four research groups},
  journal       = {IEEE Signal Process. Mag.},
  volume        = {29},
  number        = {6},
  pages         = {82--97},
  year          = {2012}
}

@inproceedings{FCN_fu2017raw,
  author        = {S.-W. Fu and Y. Tsao and X. Lu and H. Kawai},
  title         = {Raw waveform-based speech enhancement by fully convolutional networks},
  booktitle     = {Proc. APSIPA ASC},
  year          = {2017}
}

@inproceedings{pandey2020densely,
  author        = {A. Pandey and D. Wang},
  title         = {Densely connected neural network with dilated convolutions for real-time speech enhancement in the time domain},
  booktitle     = {Proc. ICASSP},
  pages         = {6629--6633},
  year          = {2020}
}

@inproceedings{RNN_valentini2016investigating,
  author        = {C. Valentini-Botinhao and X. Wang and S. Takaki and J. Yamagishi},
  title         = {Investigating {RNN}-based speech enhancement methods for noise-robust text-to-speech},
  booktitle     = {Proc. SSW},
  year          = {2016}
}

@inproceedings{LSTM_weninger2015speech,
  author        = {F. Weninger and H. Erdogan and S. Watanabe and E. Vincent and J. {Le Roux} and J. R. Hershey and B. Schuller},
  title         = {Speech enhancement with {LSTM} recurrent neural networks and its application to noise-robust {ASR}},
  booktitle     = {Proc. LVA/ICA},
  year          = {2015},
  doi           = {10.1007/978-3-319-22482-4_11}
}

@article{vaswani2017attention,
  author        = {Y.-J. Lu and C.-Y. Chang and C. Yu and C.-F. Liu and J.-W. Hung and S. Watanabe and Y. Tsao},
  title         = {Improving speech enhancement performance by leveraging contextual broad phonetic class information},
  journal       = {IEEE/ACM Trans. Audio Speech Lang. Process.},
  volume        = {31},
  pages         = {2738--2750},
  year          = {2023},
  doi           = {10.1109/TASLP.2023.3288418}
}

@inproceedings{SEMamba,
  author        = {R. Chao and W.-H. Cheng and M. {La Quatra} and S. M. Siniscalchi and C.-H. H. Yang and S.-W. Fu and Y. Tsao},
  title         = {An investigation of incorporating {Mamba} for speech enhancement},
  booktitle     = {Proc. SLT},
  pages         = {302--308},
  year          = {2024},
  doi           = {10.1109/SLT61566.2024.10832332}
}

@article{abdulatif2024cmgan,
  author        = {S. Abdulatif and R. Cao and B. Yang},
  title         = {{CMGAN}: {Conformer}-based metric-gan for monaural speech enhancement},
  journal       = {IEEE/ACM Trans. Audio Speech Lang. Process.},
  volume        = {32},
  pages         = {2477--2493},
  year          = {2024}
}

@article{wang2023tf,
  author        = {Z.-Q. Wang and S. Cornell and S. Choi and Y. Lee and B.-Y. Kim and S. Watanabe},
  title         = {{TF-GridNet}: Integrating full-and sub-band modeling for speech separation},
  journal       = {IEEE/ACM Trans. Audio Speech Lang. Process.},
  volume        = {31},
  pages         = {3221--3236},
  year          = {2023}
}

@inproceedings{fu2019metricgan,
  author        = {S.-W. Fu and C.-F. Liao and Y. Tsao and S.-D. Lin},
  title         = {{MetricGAN}: Generative adversarial networks based black-box metric scores optimization for speech enhancement},
  booktitle     = {Proc. PMLR},
  pages         = {2031--2041},
  year          = {2019}
}

@inproceedings{kolbaek2018monaural,
  author        = {M. Kolb{\ae}k and Z.-H. Tan and J. Jensen},
  title         = {Monaural speech enhancement using deep neural networks by maximizing a short-time objective intelligibility measure},
  booktitle     = {Proc. ICASSP},
  pages         = {5059--5063},
  year          = {2018}
}

@inproceedings{fu2021metricgan+,
  author        = {S.-W. Fu and C. Yu and T.-A. Hsieh and P. Plantinga and M. Ravanelli and X. Lu and Y. Tsao},
  title         = {{MetricGAN+}: An improved version of metricgan for speech enhancement},
  booktitle     = {Proc. Interspeech},
  year          = {2021},
  pages         = {201--205},
  doi           = {10.21437/Interspeech.2021-599}
}

@article{nwdrc2025,
  author        = {H. Zhang and B. C. J. Moore and F. Jiang and M. Diao and F. Ji and X. Li and C. Zheng},
  title         = {{Neural-WDRC}: A deep learning wide dynamic range compression method combined with controllable noise reduction for hearing aids},
  journal       = {Trends Hear.},
  volume        = {29},
  year          = {2025},
  doi           = {10.1177/23312165241309301}
}

@article{kayser2022spatial,
  author        = {H. Kayser and H. Hermansky and B. T. Meyer},
  title         = {Spatial speech detection for binaural hearing aids using deep phoneme classifiers},
  journal       = {Acta Acust.},
  volume        = {6},
  pages         = {25},
  year          = {2022}
}

@article{dlha1,
  author        = {F. Drakopoulos and S. Verhulst},
  title         = {A neural-network framework for the design of individualised hearing-loss compensation},
  journal       = {IEEE/ACM Trans. Audio Speech Lang. Process.},
  volume        = {31},
  pages         = {2395--2409},
  year          = {2023}
}

@inproceedings{dlha2,
  author        = {F. Drakopoulos and A. {Van Den Broucke} and S. Verhulst},
  title         = {A {DNN}-based hearing-aid strategy for real-time processing: One size fits all},
  booktitle     = {Proc. ICASSP},
  pages         = {1--5},
  year          = {2023}
}

@inproceedings{dlha3,
  author        = {F. Drakopoulos and S. Verhulst},
  title         = {A differentiable optimisation framework for the design of individualised {DNN}-based hearing-aid strategies},
  booktitle     = {Proc. ICASSP},
  pages         = {351--355},
  year          = {2022}
}

@article{dlha4,
  author        = {P. Leer and J. Jensen and Z.-H. Tan and J. {\O}stergaard and L. Bramsl{\o}w},
  title         = {How to train your ears: Auditory-model emulation for large-dynamic-range inputs and mild-to-severe hearing losses},
  journal       = {IEEE/ACM Trans. Audio Speech Lang. Process.},
  volume        = {32},
  pages         = {2006--2020},
  year          = {2024}
}

@article{ahmed2025neuroamp,
  author        = {S. Ahmed and R. E. Zezario and H.-G. Yuan and A. Hussain and H.-M. Wang and W.-H. Chung and Y. Tsao},
  title         = {{NeuroAMP}: A novel end-to-end general purpose deep neural amplifier for personalized hearing aids},
  journal       = {IEEE Trans. Artif. Intell.},
  year={2026},
  volume={7},
  number={3},
  pages={1610--1625},
}

@article{brons2014effects,
  author        = {I. Brons and R. Houben and W. A. Dreschler},
  title         = {Effects of noise reduction on speech intelligibility, perceived listening effort, and personal preference in hearing-impaired listeners},
  journal       = {Trends Hear.},
  volume        = {18},
  pages         = {2331216514553924},
  year          = {2014}
}

@article{kolbaek2016speech,
  author        = {M. Kolb{\ae}k and Z.-H. Tan and J. Jensen},
  title         = {Speech intelligibility potential of general and specialized deep neural network based speech enhancement systems},
  journal       = {IEEE/ACM Trans. Audio Speech Lang. Process.},
  volume        = {25},
  number        = {1},
  pages         = {153--167},
  year          = {2017}
}

@article{owens1985visemes,
  author  = {E. Owens and B. Blazek},
  title   = {Visemes observed by hearing-impaired and normal-hearing adult viewers},
  journal = {J. Speech Hear. Res.},
  volume  = {28},
  number  = {3},
  pages   = {381--393},
  year    = {1985},
  doi     = {10.1044/jshr.2803.381}
}

@article{vantasell1987envelope,
  author  = {D. J. {Van Tasell} and S. D. Soli and V. M. Kirby and G. P. Widin},
  title   = {Speech waveform envelope cues for consonant recognition},
  journal = {J. Acoust. Soc. Am.},
  volume  = {82},
  number  = {4},
  pages   = {1152--1161},
  year    = {1987},
  doi     = {10.1121/1.395251}
}

@article{grant1998avintegration,
  author  = {K. W. Grant and B. E. Walden and P. F. Seitz},
  title   = {Auditory-visual speech recognition by hearing-impaired subjects: Consonant recognition, sentence recognition, and auditory-visual integration},
  journal = {J. Acoust. Soc. Am.},
  volume  = {103},
  number  = {5},
  pages   = {2677--2690},
  year    = {1998},
  doi     = {10.1121/1.422788}
}

@article{bernstein2020bandimportance,
  author  = {J. G. W. Bernstein and J. H. Venezia and K. W. Grant},
  title   = {Auditory and auditory-visual frequency-band importance functions for consonant recognition},
  journal = {J. Acoust. Soc. Am.},
  volume  = {147},
  number  = {5},
  pages   = {3712--3727},
  year    = {2020},
  doi     = {10.1121/10.0001301}
}

@article{hou2018audio,
  author        = {J.-C. Hou and S.-S. Wang and Y.-H. Lai and Y. Tsao and H.-W. Chang and H.-M. Wang},
  title         = {{Audio-visual} speech enhancement using multimodal deep convolutional neural networks},
  journal       = {IEEE Trans. Emerg. Top. Comput. Intell.},
  volume        = {2},
  number        = {2},
  pages         = {117--128},
  year          = {2018}
}

@inproceedings{ephrat2018looking,
  author        = {A. Ephrat and I. Mosseri and O. Lang and T. Dekel and K. Wilson and A. Hassidim and W. T. Freeman and M. Rubinstein},
  title         = {Looking to listen at the cocktail party: A speaker-independent {audio-visual} model for speech separation},
  booktitle     = {Proc. SIGGRAPH},
  year          = {2018}
}

@article{michelsanti2021overview,
  author        = {D. Michelsanti and Z.-H. Tan and S.-X. Zhang and Y. Xu and M. Yu and D. Yu and J. Jensen},
  title         = {An overview of deep-learning-based {audio-visual} speech enhancement and separation},
  journal       = {IEEE/ACM Trans. Audio Speech Lang. Process.},
  volume        = {29},
  pages         = {1368--1396},
  year          = {2021}
}

@article{lalonde2022face,
  author        = {K. Lalonde and E. Buss and M. K. Miller and L. J. Leibold},
  title         = {Face masks impact auditory and {audiovisual} consonant recognition in children with and without hearing loss},
  journal       = {Front. Psychol.},
  volume        = {13},
  pages         = {874345},
  year          = {2022}
}

@article{TMSV,
  author        = {S.-Y. Chuang and H.-M. Wang and Y. Tsao},
  title         = {Improved lite {audio-visual} speech enhancement},
  journal       = {IEEE/ACM Trans. Audio Speech Lang. Process.},
  volume        = {30},
  pages         = {1345--1359},
  year          = {2022},
  doi           = {10.1109/TASLP.2022.3153265}
}

@inproceedings{AVTransformer2023,
  author        = {V. {Ahmadi Kalkhorani} and A. Kumar and K. Tan and B. Xu and D. Wang},
  title         = {Time-domain {Transformer}-based {audiovisual} speaker separation},
  booktitle     = {Proc. Interspeech},
  pages         = {3472--3476},
  year          = {2023},
  doi           = {10.21437/Interspeech.2023-2098}
}

@misc{SAVSE2025,
  author        = {R. Chao and W. Ren and Y.-J. Li and K.-H. Hung and S.-F. Huang and S.-W. Fu and W.-H. Cheng and Y. Tsao},
  title         = {Leveraging {Mamba} with full-face vision for {audio-visual} speech enhancement},
  howpublished  = {{arXiv} preprint},
  eprint        = {2508.13624},
  archivePrefix = {arXiv},
  year          = {2025},
  url           = {https://arxiv.org/abs/2508.13624}
}

@inproceedings{martel2023lightweight,
  author    = {H. Martel and J. Richter and K. Li and X. Hu and T. Gerkmann},
  title     = {Audio-visual speech separation in noisy environments with a lightweight iterative model},
  booktitle = {Proc. Interspeech},
  pages     = {1673--1677},
  year      = {2023},
  doi       = {10.21437/Interspeech.2023-1753}
}

@inproceedings{AVSE5,
  author        = {S.-Y. Chuang and Y. Tsao and C.-C. Lo and H.-M. Wang},
  title         = {Lite {audio-visual} speech enhancement},
  booktitle     = {Proc. Interspeech},
  year          = {2020}
}

@article{Conversion,
  author        = {M. Gogate and K. Dashtipour and A. Adeel and A. Hussain},
  title         = {{CochleaNet}: A robust language-independent {audio-visual} model for real-time speech enhancement},
  journal       = {Inf. Fusion},
  volume        = {63},
  pages         = {273--285},
  year          = {2020}
}

@inproceedings{yang2026tracking,
  author        = {H.-C. Yang and Y.-J. Li and R. Chao and Y. Tsao and B. Su and S.-Y. Chien},
  title         = {Tracking listener attention: Gaze-guided {audio-visual} speech enhancement framework},
  booktitle     = {Proc. ICASSP},
  pages         = {18932--18936},
  year          = {2026},
  doi           = {10.1109/ICASSP55912.2026.11461886}
}

@article{lesica2021ai,
  author  = {N. A. Lesica and N. Mehta and J. G. Manjaly and L. Deng and B. S. Wilson and F.-G. Zeng},
  title   = {Harnessing the power of artificial intelligence to transform hearing healthcare and research},
  journal = {Nat. Mach. Intell.},
  volume  = {3},
  pages   = {840--849},
  year    = {2021},
  doi     = {10.1038/s42256-021-00394-z}
}

@inproceedings{10023284,
  author        = {A. L. A. Blanco and C. Valentini-Botinhao and O. Klejch and M. Gogate and K. Dashtipour and A. Hussain and P. Bell},
  title         = {{AVSE} challenge: {Audio-visual} speech enhancement challenge},
  booktitle     = {Proc. SLT},
  pages         = {465--471},
  year          = {2023}
}

@inproceedings{Perez,
  author        = {E. Perez and F. Strub and H. de Vries and V. Dumoulin and A. Courville},
  title         = {{FiLM}: Visual reasoning with a general conditioning layer},
  booktitle     = {Proc. AAAI},
  volume        = {32},
  year          = {2018}
}

@inproceedings{CL,
  author        = {V. Zadorozhnyy and Q. Ye and K. Koishida},
  title         = {{SCP-GAN}: Self-correcting discriminator optimization for training consistency preserving metric GAN on speech enhancement tasks},
  booktitle     = {Proc. Interspeech},
  pages         = {2463--2467},
  year          = {2023},
  doi           = {10.21437/Interspeech.2023-456}
}

@inproceedings{graetzer2021clarity,
  author        = {S. Graetzer and J. Barker and T. J. Cox and M. Akeroyd and J. F. Culling and G. Naylor and E. Porter and R. {Viveros Mu{\~n}oz}},
  title         = {{Clarity-2021} challenges: Machine learning challenges for advancing hearing aid processing},
  booktitle     = {Proc. Interspeech},
  pages         = {686--690},
  year          = {2021},
  doi           = {10.21437/Interspeech.2021-1574}
}

@manual{ASHA2005,
  author        = {{American Speech-Language-Hearing Association}},
  title         = {Guidelines for manual pure-tone threshold audiometry},
  organization  = {{American Speech-Language-Hearing Association}},
  address       = {Rockville, MD},
  year          = {2005},
  url           = {https://www.asha.org/policy/gl2005-00014/},
  note          = {(Last viewed August 24, 2026)}
}

@misc{afouras2018lrs3,
  author        = {T. Afouras and J. S. Chung and A. Zisserman},
  title         = {{LRS3-TED}: A large-scale dataset for visual speech recognition},
  howpublished  = {{arXiv} preprint},
  eprint        = {1809.00496},
  archivePrefix = {arXiv},
  year          = {2018},
  url           = {https://arxiv.org/abs/1809.00496}
}

@article{thiemann2013diverse,
  author        = {J. Thiemann and N. Ito and E. Vincent},
  title         = {The diverse environments multi-channel acoustic noise database (demand): A database of multichannel environmental noise recordings},
  journal       = {Proc. Meet. Acoust.},
  volume        = {19},
  pages         = {035081},
  year          = {2013},
  doi           = {10.1121/1.4799597}
}

@inproceedings{reddy2021icassp,
  author        = {C. K. Reddy and H. Dubey and V. Gopal and R. Cutler and S. Braun and H. Gamper and R. Aichner and S. Srinivasan},
  title         = {{ICASSP} 2021 deep noise suppression challenge},
  booktitle     = {Proc. ICASSP},
  pages         = {6623--6627},
  year          = {2021}
}

@article{Haspi1,
  author        = {J. M. Kates and K. H. Arehart},
  title         = {The hearing-aid speech perception index ({HASPI})},
  journal       = {Speech Commun.},
  volume        = {65},
  pages         = {75--93},
  year          = {2014}
}

@article{Haspi2,
  author        = {J. M. Kates and K. H. Arehart},
  title         = {The hearing-aid speech perception index ({HASPI}) version 2},
  journal       = {Speech Commun.},
  volume        = {131},
  pages         = {35--46},
  year          = {2021}
}

@article{Hasqi1,
  author        = {J. M. Kates and K. H. Arehart},
  title         = {The hearing-aid speech quality index ({HASQI}) version 2},
  journal       = {J. Audio Eng. Soc.},
  volume        = {62},
  number        = {3},
  pages         = {99--117},
  year          = {2014}
}

@article{MSBG1,
  author        = {B. C. J. Moore and B. R. Glasberg and T. Baer},
  title         = {A model for the prediction of thresholds, loudness, and partial loudness},
  journal       = {J. Audio Eng. Soc.},
  volume        = {45},
  pages         = {224--240},
  year          = {1997}
}

@article{MSBG2,
  author        = {Y. Nejime and B. C. Moore},
  title         = {Simulation of the effect of threshold elevation and loudness recruitment combined with reduced frequency selectivity on the intelligibility of speech in noise},
  journal       = {J. Acoust. Soc. Am.},
  volume        = {102},
  number        = {1},
  pages         = {603--615},
  year          = {1997}
}

@article{brown2016binaural,
  author        = {A. D. Brown and F. A. Rodriguez and C. D. F. Portnuff and M. J. Goupell and D. J. Tollin},
  title         = {Time-varying distortions of binaural information by bilateral hearing aids: Effects of nonlinear frequency compression},
  journal       = {Trends Hear.},
  volume        = {20},
  pages         = {2331216516668303},
  year          = {2016},
  doi           = {10.1177/2331216516668303}
}

@article{peng2022spatialrelease,
  author        = {Z. E. Peng and R. Y. Litovsky},
  title         = {Novel approaches to measure spatial release from masking in children with bilateral cochlear implants},
  journal       = {Ear Hear.},
  volume        = {43},
  number        = {1},
  pages         = {101--114},
  year          = {2022},
  doi           = {10.1097/AUD.0000000000001080}
}

@inproceedings{yeh2024hyperrnn,
  author    = {Y.-T. Yeh and W.-Y. Hsiao and Y.-H. Yang},
  title     = {Hyper Recurrent Neural Network: Condition Mechanisms for
               Black-box Audio Effect Modeling},
  booktitle = {Proc. DAFx},
  year      = {2024}
}

@article{ni2026affine,
  author  = {Y. Ni and R. Liang and X. Hao and J. Cheng and Q. Wang and
             C. Huang and C. Zou and W. Zhou and W. Ding and B. W. Schuller},
  title   = {Affine Modulation-based Audiogram Fusion Network for Joint
             Noise Reduction and Hearing Loss Compensation},
  journal = {Inf. Fusion},
  volume  = {127},
  pages   = {103726},
  year    = {2026},
  doi     = {10.1016/j.inffus.2025.103726}
}

@inproceedings{peebles2023dit,
  author    = {W. Peebles and S. Xie},
  title     = {Scalable Diffusion Models with Transformers},
  booktitle = {Proc. ICCV},
  pages     = {4172--4182},
  year      = {2023},
  doi       = {10.1109/ICCV51070.2023.00387}
}

@inproceedings{birnbaum2019temporal,
  author    = {S. Birnbaum and V. Kuleshov and S. Z. Enam and P. W. Koh and S. Ermon},
  title     = {Temporal {FiLM}: Capturing long-range sequence dependencies with feature-wise modulations},
  booktitle = {Adv. Neural Inf. Process. Syst.},
  volume    = {32},
  pages     = {10287--10298},
  year      = {2019}
}

@inproceedings{MeseguerBrocal2019,
  author    = {G. Meseguer-Brocal and G. Peeters},
  title     = {{Conditioned-U-Net}: Introducing a control mechanism in the {U-Net} for multiple source separation},
  booktitle = {Proc. ISMIR},
  year      = {2019}
}

@article{chadha2021world,
  title={The world report on hearing},
  author={S. Chadha and K. Kamenov and A. Cieza},
  journal={Bull. World Health Organ.},
  volume={99},
  number={4},
  pages={242},
  year={2021}
}

@inproceedings{chao2026rtsemamba,
  title     = {{RT-SEMamba}: Real-Time Speech Enhancement Mamba via Progressive Knowledge Distillation},
  author    = {R. Chao and S.-F. Huang and M. {La Quatra} and S. M. Siniscalchi and W.-H. Cheng and S.-W. Fu and Y. Tsao},
  booktitle = {Proc. Interspeech},
  year      = {2026}
}

@article{ni2026fpga,
  author  = {X. Ni and D. Xu and U. Anwar and T. Arslan},
  title   = {A Real-Time {FPGA}-Accelerated Causal Spectro-Temporal Reconstruction Network for Speech Enhancement in Hearing Aids},
  journal = {IEEE Trans. Very Large Scale Integr. (VLSI) Syst.},
  pages   = {1--12},
  year    = {2026},
  doi     = {10.1109/TVLSI.2026.3716946}
}

\end{document}